\documentclass[12pt]{article}

\usepackage{graphicx}
\usepackage{amssymb}
\usepackage{amsmath}
\usepackage{epsf}
\usepackage[parfill]{parskip}
\usepackage[matrix,arrow]{xy}
\usepackage[usenames]{color}
\usepackage{mathrsfs}
\input{epsf}

\newcommand{\ba}{\begin{eqnarray*}}
\newcommand{\ea}{\end{eqnarray*}}
\newcommand{\ban}{\begin{eqnarray}}
\newcommand{\ean}{\end{eqnarray}}

\newcommand{\rank}{{\rm rank}}

\newcommand{\ch}{{\rm ch}}

\newcommand{\IZ}{\mathbb{Z}}
\newcommand{\IC}{\mathbb{C}}
\newcommand{\IP}{\mathbb{P}}

\newcommand{\IQ}{\mathbb{Q}}

\newcommand{\cN}{{\cal N}}

\newcommand{\cP}{{\cal P}}

\newcommand{\txi}{\tilde{\xi}}
\newcommand{\tk}{\tilde{k}}

\newcommand{\tomega}{\tilde{\omega}}

\newcommand{\tP}{\tilde{\cal P}}

\newcommand{\re}{{\rm Re \,}}
\newcommand{\im}{{\rm Im \,}}

\def \nn{\nonumber}

\makeatletter
\@addtoreset{equation}{section}
\makeatother

\newcommand{\be}{\begin{equation}}
\newcommand{\ee}{\end{equation}}
\newcommand{\bea}{\begin{eqnarray}}
\newcommand{\eea}{\end{eqnarray}}
\newcommand{\nnb}{\nonumber}

\newcommand{\om}{\omega}
\newcommand{\Om}{\Omega}

\newcommand{\cl}[1]{\mathcal{#1}}
\newcommand{\ul}{\underline}

\begin{document}

\renewcommand{\thefootnote}{\fnsymbol{footnote}}
\setcounter{footnote}{0}

\begin{titlepage}
\begin{flushright}
IHES/P/09/03\\
LPTENS-09/02\\
ROM2F/2009/01\\
January 2009
\end{flushright}
\begin{center}
\vskip 2cm {\Huge Exploiting $\cl N =2$ in consistent coset reductions of type IIA 
\\ \vskip 0.1cm}
\vskip 1cm {Davide Cassani$^{a,b}$\footnote{cassani AT lpt.ens.fr} and Amir-Kian Kashani-Poor$^{\,c}$\footnote{kashani AT ihes.fr}}
\vskip.6cm 
{\it $^a$ Laboratoire de Physique Th\'eorique\footnote{Unit\'e mixte du CNRS et
    de l'\'Ecole Normale Sup\'erieure associ\'ee \`a l'Universit\'e Pierre et
    Marie Curie Paris 6, UMR
    8549.}$\!$,
\'Ecole Normale Sup\'erieure,
\\
24 rue Lhomond, 75231 Paris Cedex 05, France\\
\vskip .3cm
$^b$ Dipartimento di Fisica, Universit\`a di Roma ``Tor Vergata''\\
Via della Ricerca Scientifica, 00133 Roma, Italy\\
\vskip .3cm 
$^c$ Institut des Hautes \'Etudes Scientifiques\\
Le Bois-Marie, 35, route de Chartres \\
91440 Bures-sur-Yvette, France\\ \vskip0.2cm }
\end{center} 
\vskip 1.5cm
\begin{abstract}
We study compactifications of type IIA supergravity on cosets exhibiting SU(3) structure. We establish the consistency of the truncation based on left-invariance, providing a justification for the choice of expansion forms which yields gauged $\cN=2$ supergravity in four dimensions. We explore $\cN=1$ solutions of these theories, emphasizing the requirements of flux quantization, as well as their non-supersymmetric companions. In particular, we obtain a no-go result for de Sitter solutions at string tree level, and, exploiting the enhanced leverage of the $\cN=2$ setup, provide a preliminary analysis of the existence of de Sitter vacua at all string loop order. 
\end{abstract}

\end{titlepage}
\newpage


\renewcommand{\thefootnote}{\arabic{footnote}}
\setcounter{footnote}{0}

\section{Introduction}
In the era of LHC, much effort is being invested in finding phenomenologically viable string vacua. Much of this work takes place by considering compactifications to $\cN=1$ theories in 4d. In this paper, we will focus instead on a framework which yields 4-dimensional theories that have $\cN=2$ symmetry realized off-shell. While the $\cN=1$ setup allows for more flexibility in choosing the various ingredients of the theory, and hence (currently) permits the construction of more realistic vacua, the increased rigidity of the $\cN=2$ setup has the advantage of allowing a more exhaustive treatment of $\alpha'$, string loop, and foreseeably even brane instanton corrections. An impressive example of the power of the $\cN=2$ framework is the recent proof \cite{Louis} that $\cN=2$ gauged supergravities without vector multiplets do not permit de Sitter vacua, in spite of the presence of such solutions in the one-brane-instanton approximation \cite{Vandoren}. Studying theories in the $\cN=2$ framework hence presents one promising avenue towards assessing the viability of the approximations that are necessary to get off the ground in less supersymmetric frameworks.

The best studied example of $\cN=2$ theories obtained from string theory are type II Calabi-Yau compactifications \cite{BodnerCadavidFerrara, BohmGHLouis}. The differential operators governing the geometric moduli problem of the internal Calabi-Yau manifolds turn out to coincide with the mass operators of the supergravity theory. Unobstructed deformations hence give rise to massless excitations, resulting in the beautiful identification between the massless scalar fields of these theories, whose VEVs parametrize a family of supergravity solutions, and the geometric moduli of the Calabi-Yau. The masslessness of the scalars is protected by supersymmetry, as $\cN=2$ forbids a potential in the case of uncharged matter. In \cite{generalized mirror symmetry}, the study of type II compactifications on SU(3) structure manifolds was initiated (recall that Calabi-Yau manifolds satisfy the stronger condition of SU(3) holonomy). This setup is more akin to the phenomenologically motivated $\cN=1$ analyses: solutions of the supergravity equations of motion on these internal manifolds require the presence of background fluxes \cite{GiddingsKachruPolchinski, GMPT1, GMPT2}, and compactification gives rise to 4d $\cN=2$ gauged supergravity theories \cite{PolchinskiStrominger, N=2review}, which, in contrast to the Calabi-Yau case with uncharged matter, exhibit a potential for the scalar fields in the theory. The increased phenomenological viability comes at a price: the very presence of a potential makes it unlikely that the choice of light degrees of freedom of the theory can be associated to a geometric moduli problem. Indeed, a systematic approach to a reduction ansatz for these theories is still lacking. Following our work in \cite{ReducingSU3SU3} and \cite{KashaniPoorNearlyKahler}, we here pursue an alternative approach towards justifying the reduction ansatz, that of consistent truncation: obtaining a field theory with a finite number of fields upon compactification requires truncating most of the degrees of freedom of the higher dimensional theory; this truncation is called consistent when all solutions to the lower dimensional equations of motion lift to solutions of the higher dimensional theory. Note the contrast to a Kaluza-Klein reduction \cite{DuffNilssonPopeKKreview}, which is an expansion valid around a single 10d solution (hence referred to as a base-point dependent reduction in \cite{KashaniPoorMinasian}). 

Consistently truncated lower dimensional field theories are powerful allies in studying the vacuum structure of the higher dimensional {\it string} theory. This is partially a consequence of computational techniques being more refined in lower dimensions. E.g., various leading non-trivial contributions in $\alpha'$ to the 10d type II supergravity action have been determined \cite{GrossWitten,GreenSchwarz,SakaiTanii,PolicastroTsimpis}. One may hope to establish the complete action to this order by 10d supersymmetric completion \cite{PeetersVW}. However, the 10d supersymmetry equations have simply proved too cumbersome to date. By contrast, the supersymmetric completion of the contribution of these terms to the 4d $\cl N=2$ supergravity action is readily available, yielding the full string tree level and one loop corrected action. In fact, in 4d we can, as we will discuss, even draw conclusions regarding the all string loop corrected action. Studying the lower dimensional theory is however not merely a question of computational convenience. An effective higher dimensional description of worldsheet or brane instantons is even conceptually problematic.

In \cite{KashaniPoorNearlyKahler}, it was shown that expansion forms can be defined on Nearly K\"ahler manifolds that satisfy the conditions of \cite{KashaniPoorMinasian}, implying that the reduction of the type IIA action based on these forms yields $\cN=2$ gauged supergravity in 4d. It was further demonstrated that the truncation in this setting is consistent in the supersymmetric sector (i.e. 4d solutions preserving $\cN=1$ supersymmetry lift). In this paper, we shift our focus to certain coset spaces which subsume the currently known set of 6d Nearly K\"ahler manifolds. We introduce these spaces in section \ref{introinter}. Considering the emphasis on base point independence of the reduction, it was perhaps somewhat disappointing that the theories based on Nearly K\"ahler reduction yielded a single supersymmetric vacuum for a given choice of fluxes. Cosets by contrast permit multiple $\cN=1$ solutions for a given choice, which are all accessible via the 4d theory. We demonstrate this in section \ref{sussol}. Due to flux quantization, the solutions come in a discrete family. We perform the required $K$-theory analysis. In section \ref{consistent}, we demonstrate that the left-invariant coset reductions represent a consistent truncation by establishing that the 10d equations of motion reduce to the 4d equations following from the appropriate $\cl N=2$ action. This extends the analysis of \cite{ReducingSU3SU3} beyond the RR sector and overcomes the restriction to consistency merely of the supersymmetric sector \cite{KashaniPoorNearlyKahler, CassaniBilal}. Fueled by this result, we turn to the study of non-supersymmetric vacua of the 4d theories in sections \ref{4dpotential} and \ref{nonsusy}. We find several non-supersymmetric Nearly K\"ahler companions to the solution of section \ref{sussol} and study their stability, in particular with regard to deformations away from the Nearly K\"ahler locus. We also consider the question of the existence of de Sitter vacua, which has received some attention recently in the type IIA context \cite{Hertzberg,MinimalSimpledS, Louis, CosmologySU3, FlaugerPabanRobbinsWrase, Neupane}. We demonstrate that such vacua are absent at string tree level (we prove this result in greater generality than the coset context: it is valid for any gauged supergravity with merely the universal tree-level hypermultiplet, irrespective of the specifics of the vector multiplet sector). Due to the increased leverage in the $\cN=2$ setup, we are able to push this analysis beyond tree level. We obtain the full string loop corrected potential, which evades the tree-level no-go theorem, and uncover a necessary condition on the contribution of the NSNS sector to the potential for de Sitter vacua to be possible. In two appendices, we fill in the details of the dimensional reduction leading to the 4d $\cN=2$ theory (appendix \ref{DetailsDimRed}), and study the string loop corrected 4d $\cN=1$ conditions (appendix \ref{LoopCorrectionsToN=1}).

\section{Introducing the internal geometries} \label{introinter}

We consider dimensional reductions of massive type IIA supergravity on left coset spaces $M_6=G/H$ endowed with a left-invariant SU(3) structure. An exhaustive list of such cosets was provided in ref.$\:$\cite{KoerberLustTsimpis} (see section 1 and in particular table 1 therein). In the following, we are going to focus on the cosets whose SU(3) structure cannot be further reduced to SU(2), namely 
\be\label{eq:OurCosets}
\frac{\mathrm{SU(3)}}{\mathrm{U(1)}\times \mathrm{U(1)}}\quad, \quad\frac{\mathrm{Sp(2)}}{\mathrm{S(U(2)}\times\mathrm{U(1))}}\quad,\quad\frac{\mathrm{G}_2}{\mathrm{SU(3)}}\;,
\ee
where $\mathrm{S(U(2)}\times\mathrm{U(1))}$ is non-maximally embedded in Sp(2). 

It is easy to see that a reduction performed on these manifolds by expanding the higher dimensional fields in a basis of left-invariant forms satisfies the constraints of \cite{KashaniPoorMinasian} and therefore yields a gauged $\cN=2$ supergravity in 4d. 

The remaining cosets listed in \cite{KoerberLustTsimpis} have vanishing Euler characteristic and admit a left-invariant vector: their SU(3) structure group is therefore further reduced to at least SU(2). For these cosets, the $\cN=2$ reduction ansatz based on the presence of SU(3) structure can be more naturally enlarged to include the whole set of left-invariant forms, possibly yielding a further extended supergravity ($\cl N\geq 4$) in 4d.

The only non-vanishing torsion classes\footnote{For a review of SU(3) structures and their torsion classes, see e.g. subsection 3.2 of ref. \cite{GranaReview}.} characterizing the SU(3) structure of the cosets (\ref{eq:OurCosets}) are $W_1$ and $W_2$, i.e. the SU(3) invariant 2- and 3-form $J$ and $\Om$ satisfy
\bea
\nnb dJ &=& \frac{3}{2}\im(\bar W_1 \Om) \;, \\ 
\label{eq:TorsionClasses} d \Om &=& W_1 J\wedge J + W_2\wedge J\;.
\eea
In fact, $\frac{\textrm{G}_2}{\textrm{SU(3)}}$ allows just $W_1\neq 0$ and is therefore a Nearly K\"ahler manifold. The cosets $\frac{\textrm{SU(3)}}{\textrm{U(1)}\times \textrm{U(1)}}$ and $\frac{\textrm{Sp(2)}}{\textrm{S(U(2)}\times\textrm{U(1))}}$ also admit a region in the SU(3) structure parameter space in which they are Nearly K\"ahler, but in general, their $W_2$ torsion class does not vanish. Since $W_1$ and $W_2$ can be chosen purely imaginary, these cosets fall into the class of `half-flat' manifolds, characterized by $\re W_1 = \re W_2 = W_4 = W_5 = 0$ \cite{ChiossiSalamon}.

A description of the coset spaces (\ref{eq:OurCosets}) was given e.g. in \cite{MuellerStuckl}. In the context of SU(3) structure compactifications of (massive) type IIA supergravity, supersymmetric AdS$_4$ backgrounds on these manifolds have recently been found in \cite{KoerberLustTsimpis, BehrndtCveticShort, HousePalti, TomasielloTwistor} and further discussed in \cite{AldazabalFont}, while refs.$\:$\cite{CaviezelKoerberKorsLustTsimpisZagermann, CosmologySU3} study the properties of the associated effective 4d $\cl N=1$ supergravity in the presence of orientifold projections (see also \cite{HousePalti} for a previous work considering the coset $\frac{\mathrm{SU(3)}}{\mathrm{U(1)}\times \mathrm{U(1)}}$). Type IIA reduction on Nearly K\"ahler manifolds has been worked out in \cite{KashaniPoorNearlyKahler}. The cosets (\ref{eq:OurCosets}) appeared in the string literature in \cite{Lust1,CastellaniLust} in the heterotic context, and have also been employed recently in \cite{HeteroticOnCosets} for heterotic dimensional reductions.

\subsection{The expansion forms}\label{ExpBasis}

In the following we provide the most general left-invariant positive-definite metric for each coset (\ref{eq:OurCosets}), as well as a basis for all the left-invariant differential forms, on which we are going to expand the supergravity fields.

We define the 6d coset spaces (\ref{eq:OurCosets}) as in ref.$\;$\cite{KoerberLustTsimpis}, and in particular adopt the set of group structure constants listed therein. The same reference also provides a summary of the needed mathematical notions about coset spaces, while a more extended review can be found e.g. in \cite{MuellerStuckl}.

Using the local coframe\footnote{Here and in the following (see in particular subsection \ref{Consistency}), frame indices are underlined.} $\{e^{\ul m}\}$ inherited from $G$, a differential form on the coset $G/H$ reads $\omega_k=\frac{1}{k!}\omega_{\ul m_1\ldots,\ul m_k}e^{\ul m_1}\wedge\dots \wedge e^{\ul m_k}$. This is invariant under the left action of $G$ if its components are constant and satisfy the following relation involving the $G$ structure constants 
\be\label{eq:LeftInv}
f^{\ul p}{}_{i [ \ul m_1}\, \omega_{\ul m_2\ldots\ul m_k ]\ul p} = 0\;,
\ee
where the index $i$ is associated with the generators of the algebra $\mathfrak h$, while the underlined indices label a basis for the complement of $\mathfrak h$ in $\mathfrak g$. For the coset metric $ds^2 = g_{\ul{mn}}e^{\ul m}\otimes e^{\ul n}$ the relation is analogous to (\ref{eq:LeftInv}), with a symmetrization of indices replacing the antisymmetrization. The action of the exterior derivative preserves left-invariance, and is also determined by the structure constants of $G$.

None of the cosets we consider admits left-invariant 1-- or 5--forms. 

We define the `standard volume' of the cosets as 
\be \nnb I:=\int e^{123456}\;.\ee

\subsubsection{$\frac{\textrm{SU(3)}}{\textrm{U(1)}\times \textrm{U(1)}}$}
Left-invariant metric:\vskip -5mm
\be\label{eq:LeftInvMetricSU3modU1U1}
 g_{\ul{mn}} = \mathrm{diag}(v^1,v^1,v^2,v^2,v^3,v^3)\;,\quad v^1>0,\,v^2>0,\,v^3>0\;.
\ee
The left-invariant forms are spanned by
\bea
\nnb \om_0 =1\qquad,\qquad \om_1 = -e^{12}\quad &,&\quad \om_2 = e^{34} \qquad,\qquad \om_3 = -e^{56}\;,\\ [2mm]
\nnb\alpha = \frac{1}{2\sqrt I}(e^{135} + e^{146} - e^{236} + e^{245}) \quad &,&\quad \beta = \frac{1}{2\sqrt I}( -e^{136} + e^{145} - e^{235} - e^{246}) \;,\\ [2mm]
\label{eq:BasisFormsSU3modU1U1}\tilde\om^0 =  \frac{1}{I} e^{123456}\quad,\quad\tilde\om^1 =  \frac{1}{I} e^{3456}\;\; &,&\;\; \tilde\om^2 = -\frac{1}{I} e^{1256} \quad,\quad \tilde\om^3 =  \frac{1}{I} e^{1234}\;.
\eea

\pagebreak
\subsubsection{$\frac{\mathrm{Sp(2)}}{\mathrm{S(U(2)}\times\mathrm{U(1))}}$}

Left-invariant metric:\vskip -5mm
\be\label{eq:LeftInvMetricSp2modSU2xU1}
g_{\ul{mn}} = \mathrm{diag}(v^1,v^1,v^1,v^1,v^2,v^2)\;,\qquad v^1>0,\,v^2>0\;.
\ee
Basis of left-invariant forms:
\bea
\nnb \om_0 =1\qquad ,\qquad \om_1 = -e^{12}- e^{34}\quad &,&\quad \om_2 = e^{56}\;,\\ [2mm]
\nnb\alpha = \frac{1}{2\sqrt I}(e^{135} + e^{146} + e^{236} - e^{245}) \quad &,&\quad \beta = \frac{1}{2\sqrt I}( e^{136} - e^{145} - e^{235} - e^{246}) \;,\\ [2mm]
\label{eq:BasisFormsSp2modSU2xU1}\tilde\om^0 = \frac{1}{I} e^{123456}\quad\;\; , \qquad\tilde\om^1 = \frac{1}{2I}( e^{1256} \!\!\!\!&+&\!\!\! e^{3456})\quad\; , \;\quad \tilde\om^2 =  -\frac{1}{I} e^{1234}\;.
\eea

\subsubsection{$\frac{\textrm{G}_2}{\textrm{SU(3)} }$}
Left-invariant metric:\vskip -5mm
\be\label{eq:LeftInvMetricG2modSU3}
g_{\ul{mn}} = \mathrm{diag}(v^1,v^1,v^1,v^1,v^1,v^1)\;,\qquad v^1>0\;.
\ee
Basis of left-invariant forms:\vskip -8mm
\bea
\nnb \om_0 =1\quad &,&\quad \om_1 = -e^{12}+ e^{34}-e^{56}\;,\\ [2mm]
\nnb\alpha = \frac{1}{2\sqrt I}(e^{135} + e^{146} - e^{236} + e^{245}) \quad &,&\quad \beta = \frac{1}{2\sqrt I}( -e^{136} + e^{145} - e^{235} - e^{246}) \;,\\ [2mm]
\label{eq:BasisFormsG2modSU3}\tilde\om^0 =  \frac{1}{I} e^{123456}\quad &,& \quad\tilde\om^1 =  \frac{1}{3I} (e^{3456} - e^{1256} + e^{1234})\;.
\eea

\subsubsection{Properties}\label{properties}

The overall factors in the basis forms (\ref{eq:BasisFormsSU3modU1U1}), (\ref{eq:BasisFormsSp2modSU2xU1}), and (\ref{eq:BasisFormsG2modSU3}) have been chosen in such a way that
\be\label{eq:PairingBasisForms}
\int \langle\om_A , \tilde\om^B\rangle =  \delta_A^B \qquad,\qquad \int \alpha \wedge\beta = 1 \;,
\ee
where $A=(0,a)\,,\,B=(0,b)$ and $a,b$ label the left-invariant 2-- and 4--forms. The antisymmetric pairing $\langle\,,\,\rangle$ is defined on even forms $\rho,\sigma$ as $\langle \rho,\sigma\rangle\, =\, [\lambda(\rho)\wedge \sigma]_{\mathrm{top}}$, with $\,\lambda(\rho_{k}) = (-)^{\frac{k}{2}}\rho_{k}\,$, $\,k$ being the degree of $\rho$.

The basis forms define a closed differential system,\vskip -8mm
\ban
d\om_a &=& q_a\alpha \,, \nn \\
  d\alpha = 0 \; &,& \; d\beta = q_a\tilde\om^a \,,\nn \\
 d\tilde\om^A &=&0\,, \label{diffrel}
\ean
which is also closed under the action of the Hodge star operator,
\ba
*\alpha = \beta   \quad , \quad *\tilde\om^0 = \frac{1}{V\!ol}   \; &,& \;  *\tilde\om^a =-\frac{1}{4V\!ol}\cl G^{ab}\om_b \,.
\ea
Here, the $q_a$ encode what are sometimes referred to as geometric fluxes, $V\!ol$ denotes the volume of the coset, and the matrix $\cl G^{ab}$ is the inverse of 
\be\label{eq:Gfrom2forms}
\cl G_{ab} = \frac{1}{4V\!ol}\int\om_a\wedge * \om_b\;,
\ee
corresponding to the special K\"ahler metric on the space of the internal metric and B-field deformations \cite{KashaniPoorMinasian}; see subsection$\:$\ref{SpKahlerGeo} of the appendix for more details. 

In table \ref{DetailsCosets}, we give the values of the quantities introduced above for each coset.
\begin{table}
\begin{center}
\begin{tabular}{r|c|c|c}
              & $\frac{\mathrm{SU(3)}}{\mathrm{U(1)}\times \mathrm{U(1)}}$ & $\frac{\mathrm{Sp(2)}}{\mathrm{S(U(2)}\times\mathrm{U(1))}}$  &  $\frac{\mathrm{G}_2}{\mathrm{SU(3)}}$	 \\
\hline
\rule{0pt}{2.5ex} range of $a\;:\;$ &  $1,2,3$   & $1,2$  & $1$ \\ [2mm]
\rule{0pt}{0ex}   geometric flux $q_a\;:\;$ & $q_1 = q_2 = q_3 = -\sqrt I$    &  $q_1=2\sqrt I \;,\;q_2=\sqrt I $      & $q_1 = 2\sqrt{3I}$  \\ [3mm]
\rule{0pt}{0ex} $\cl G^{ab}\;=\;$ & $\mathrm{diag}\Big(\,4(v^1)^{2}\,,\,4(v^2)^{2}\,,\,4(v^3)^{2}\, \Big)$  & $\mathrm{diag}\Big(\,2(v^1)^{2}\,,\,4(v^2)^{2}\,\Big)$  &  $\frac{4}{3}(v^1)^2 $ \\ [3mm]
\rule{0pt}{0ex} $V\!ol=\;$ &  $ v^1v^2v^3I $   & $(v^1)^2v^2 I$    &  $(v^1)^3I $ \\ [2mm]
\rule{0pt}{0ex} $I=\;$ &  $  2^5\pi^3 $   & $\frac{2^7\pi^3}{3}$    &  $\frac{144\pi^3}{5} $ \\ [2mm]
\end{tabular}
\caption{Values of the different quantities introduced in subsection \ref{ExpBasis}.} \label{DetailsCosets}
\end{center}
\end{table}
The standard volume $I$ was computed following ref.$\:$\cite{MuellerStuckl}.\footnote{We have a $2^6$ supplementary factor in $I$ with respect to $\:$\cite{MuellerStuckl}. This is due to the fact that for the normalization of the group structure constants we follow the choice of \cite{KoerberLustTsimpis}, and this differs from the one of \cite{MuellerStuckl} by a factor 1/2.\label{ftn:FtnStrCnst}} Its evaluation requires knowledge of the Euler characteristic of our cosets. Since the harmonic forms on a compact coset reside among the left-invariant forms, we can read off the cohomology from the differential relations (\ref{diffrel}). We immediately conclude that all our cosets have trivial odd cohomology. Concerning the even cohomology, for $\frac{\textrm{SU(3)}}{\textrm{U(1)}\times \textrm{U(1)}}$, with
\begin{equation}
\omega'_1 = \omega_1 - \omega_3 \,, \quad \omega'_2 = \omega_2 - \omega_3 \,, \label{closed2}
\end{equation}
we have $$H^2 = {\rm Span}\left( [\omega'_1], [\omega'_2] \right) \,, \quad H^4 = {\rm Span} \left( [\tomega^1], [\tomega^2] \right) \,,$$ hence the Euler characteristic is $\chi = 6$.

For $\frac{\mathrm{Sp(2)}}{\mathrm{S(U(2)}\times\mathrm{U(1))}}$, we have $b_2=1$ and $\chi = 4$, while for $\frac{\textrm{G}_2}{\textrm{SU(3)} }$, $b_2=0$ and $\chi=2$.

\subsection{The SU(3) structure}\label{SU3str}
For each coset in (\ref{eq:OurCosets}), the pair of left-invariant forms parametrized by $v^a$,
\be\label{eq:JandOmForCoset}
J=v^a\om_a \qquad,\qquad \Om = 2\sqrt{V\!ol}(\alpha + i \beta)\,,
\ee
satisfies the relations $J\wedge \Om =0$ and $\frac{3i}{4}\Om\wedge\bar\Om =J\wedge J\wedge J$ and hence determines a left-invariant SU(3) structure. The metric specified by $J$ and $\Om$ is precisely the one given in eq.$\:$(\ref{eq:LeftInvMetricSU3modU1U1}), (\ref{eq:LeftInvMetricSp2modSU2xU1}), and (\ref{eq:LeftInvMetricG2modSU3}) respectively for the three cosets. Using the properties of the basis forms listed in subsection \ref{properties} above, one can see that the differential relations (\ref{eq:TorsionClasses}) are satisfied, with torsion classes\footnote{The evaluation of $W_2$ is performed rewriting the second line of (\ref{eq:TorsionClasses}) as $W_2=2W_1 J - *d\Om$.}
\bea\label{eq:TorsionClassesExplicit}
W_1 &=& -\frac{i v^a q_a}{3\sqrt{V\!ol}}\;, \\
\nnb W_2 &=& -\frac{2i}{3\sqrt{V\!ol}}q_a \big( v^a v^b - \frac{3}{4}\cl G^{ab}  \big)\om_b.
\eea
Substituting the quantities given in the table of subsection \ref{properties}, we see that the Nearly K\"ahler condition $W_2=0$ is identically satisfied on $\frac{\mathrm{G}_2}{\mathrm{SU(3)}}$. For $\frac{\mathrm{Sp(2)}}{\mathrm{S(U(2)}\times\mathrm{U(1))}}$ and $\frac{\textrm{SU(3)}}{\textrm{U(1)}\times \textrm{U(1)}}$, this condition is satisfied on a line in the parameter space determined by $v^1=v^2$ and $v^1=v^2=v^3$ respectively. In this Nearly K\"ahler limit the cosets are Einstein manifolds (the only other loci at which the Einstein condition is satisfied are  $2v^1=v^2$ for $\frac{\mathrm{Sp(2)}}{\mathrm{S(U(2)}\times\mathrm{U(1))}}$ and $2v^1 = 2v^2 = v^3$, or cyclic permutations of this, for $\frac{\textrm{SU(3)}}{\textrm{U(1)}\times \textrm{U(1)}}$ \cite{MuellerStuckl} ).

The forms (\ref{eq:JandOmForCoset}) are the most general left-invariant pair satisfying the SU(3) structure defining relations (the overall phase of $\Omega$ is unphysical; requiring the torsion classes to be purely imaginary, as we have done, fixes it up to a sign). In particular, since the volume $V\!ol$ is fixed by the $v^a$, we see that $\Om$ identifies a rigid SL(3,$\mathbb{C}$) structure, and there are no almost complex structure moduli.

\subsection{An alternative basis?}\label{AlternativeBasis}

In \cite{KashaniPoorMinasian}, conditions on the expansion forms were emphasized that arise when these are moduli dependent, as is the case with the basis of harmonic forms on which Calabi-Yau reductions are based (the *-ed conditions in section 2 of \cite{KashaniPoorMinasian}). For the set of expansion forms that we have introduced above, these conditions are trivially satisfied, as the forms are moduli independent. In this sense, our expansion ansatz here is technically simpler than in the Calabi-Yau case. However, in a small flux approximation, the Laplacian $\Delta= -*d*d - d*d*$ becomes the mass operator for the modes of the 10d supergravity fields, and an expansion in eigenforms of it is physically motivated. Can we replace the forms introduced above by such a basis of eigenforms?

In the Nearly K\"ahler case the expansion in eigenforms of the Laplacian is further motivated by the fact that both $J$ and $\Om$ are themselves eigenforms of $\Delta$ \cite{KashaniPoorNearlyKahler}. In the more general case $W_2\neq 0$, this is still true for $\Om$,\footnote{One needs the relation $dW_2 = \frac{i}{4} (W_2\lrcorner \bar W_2) \re\Om $, satisfied by the cosets (\ref{eq:OurCosets}).}
\be\label{eq:LaplacianOnOmega}
\Delta\Om = \big(3|W_1|^2 + \frac{1}{4}W_2\lrcorner\bar{W}_2 \big)\Om \;,
\ee
but not for $J$, which instead satisfies
\be\nnb
\Delta J = 3|W_1|^2 J - \frac{3}{2} \re(\bar W_1W_2)\;.
\ee
Considering e.g. the coset $\frac{\textrm{SU(3)}}{\textrm{U(1)}\times \textrm{U(1)}}$, a change of basis sending the 2--forms introduced in (\ref{eq:BasisFormsSU3modU1U1}) to a set of eigenforms of the Laplacian is 
\be
\om_1'=\om_1-\om_3\quad,\quad \om_2'=\om_2-\om_3\quad,\quad \om_3'= \frac{\sum_a (v^a)^2\om_a}{\sum_b(v^b)^2}\;,
\ee
where $\Delta \om_1' =\Delta\om_2'=0\,$, while $\Delta \om_3'= \frac{(v^1)^2+ (v^2)^2+ (v^3)^2}{v^1v^2v^3}\om_3'\,$. The harmonic 4--forms are spanned by
\be
*\om_1'\,\propto\, \frac{v^3}{v^1}\tilde\om^1-\frac{v^1}{v^3}\tilde\om^3\qquad,\qquad *\om_2'\,\propto\, \frac{v^3}{v^2}\tilde\om^2-\frac{v^2}{v^3}\tilde\om^3\;,
\ee
while $*\om_3'\propto -\sqrt{I}(\tilde\om^1 + \tilde\om^2 + \tilde\om^3) =  d\beta$ is exact.

The condition $v^a \partial_{v^b} \omega_a$ (*7 of \cite{KashaniPoorMinasian}) gives rise to a complicated set of equations for possible $v^a$ dependent normalization factors of the primed basis. However, it is easy to see upon inspection that the moduli independence of the triple intersection product (condition *8 of \cite{KashaniPoorMinasian}) cannot be satisfied for any such choice. The question whether the choice of left-invariant expansion forms can be motivated from a Kaluza-Klein reduction point of view hence remains an interesting open question.

\section{Supersymmetric 10d solutions parametrized by fluxes} \label{sussol}
In this section, we will rewrite the family of $\cN=1$ solutions of the 10d supergravity equations found in \cite{KoerberLustTsimpis} in a manner which makes the discreteness of this family as a result of flux quantization manifest. By \cite{CassaniBilal} and \cite{KashaniPoorNearlyKahler}, these solutions can be recovered from the 4d point of view. After proving the full consistency of our reduction in section \ref{consistent}, we will proceed to complement these solutions with their non-supersymmetric relatives in section \ref{nonsusy}.

\subsection{Flux quantization and $K$-theory}
RR-fields are classified topologically by $K$-theory classes \cite{MooreWitten,MinasianMoore}. This has two consequences for the choice of fluxes associated to the RR-fieldstrengths. Firstly, the naive integer quantization of fluxes must be replaced by quantization in multiples of fractions determined also by the topology of the compactification manifold. Secondly, not every choice of flux number satisfying these quantization conditions will possess a $K$-theory lift and hence be permissible. We will now study these two points in turn.

In \cite{MooreWitten}, fluxes were conjectured to take values in the image of the map
\ba
\sqrt{\hat{A}(X)}\, \ch(\cdot) : K(X) \rightarrow H^{{\rm even}} (X, \IQ) \,. \label{AH}
\ea 
$\ch(x)$ is the Chern character as extended to a $K$-theory element $x=E-F$ via $\ch(x) = \ch(E) - \ch(F)$.
Hence,
\ban
\frac{[F(x)]}{2 \pi} &=& \sqrt{\hat{A}}\, \ch(x)  \,,  \label{fq}
\ean
where $F = \sum_{i=0}^{5} F_{2i}$ denotes a formal sum of all RR-fieldstrengths, and $[\cdot]$ indicates rational cohomology class (rational rather than integral due to the fractional coefficients of Chern classes that appear in the expansion of the Chern character). When $H\neq 0$, the equations of motion and Bianchi identity of $F$ are modified from the naive Maxwell form, enforcing harmonicity of $F$, to a version of these equations twisted by $H$. In particular, $F$ now satisfies $(d-H)F = 0$. When $H$ is exact, as will be the case in our study, $H$-twisted cohomology maps to ordinary cohomology via $F \rightarrow e^{-B} F$, where $H = dB$. It hence proves convenient to introduce a basis of RR fields given by $G=e^{-B} F$. Equation~(\ref{fq}) then holds for $G$ rather than $F$, and the term `fluxes' refers to the cohomology classes~$[G]$. 

To decide which fluxes we can choose as boundary conditions of our physical system (and then parametrize our solutions by this choice), we need to decide on electric vs. magnetic variables. Ignoring subtleties related to torsion, which does not enter in a supergravity analysis, we can choose the electric basis to lie in $\oplus_{i=1}^{3} H^{2i}(X,\IQ)$.

Let us now consider the question of flux quantization. To this end, we expand the right hand side of (\ref{fq}) in terms of Chern classes for $x$ the class of a vector bundle $F$ on $X$,
\begin{gather*}
\ch_0(F) = \rank(F) \,, \hspace{1cm} \ch_1(F) = c_1(F) \,, \hspace{1cm} \ch_2(F) = \frac{1}{2} [c_1(F)^2 - 2c_2(F)] \,, \\
\ch_3(F) = \frac{1}{3!} [c_1(F)^3 - 3 c_1(F) c_2(F) + 3 c_3(F)] \,,\\
\hat{A} = 1 - \frac{p_1}{24} + \ldots \,.
\end{gather*}
Hence,\footnote{We thank M. Haack and L. Martucci for pointing out a term that was missing in the following formulae in a previous version of this paper.}
\begin{gather*}
\frac{[G_0]}{2\pi} = \rank(F)  \,, \hspace{1cm} \frac{[G_2]}{2\pi} = c_1(F) \,, \hspace{1cm} \frac{[G_4]}{2\pi} = \frac{1}{2} [c_1(F)^2 - 2c_2(F)]- \frac{p_1(X)}{48} \rank(F) \,, \\
\frac{[G_6]}{2\pi} = \frac{1}{3!} [c_1(F)^3 - 3 c_1(F) c_2(F) + 3 c_3(F)] - \frac{p_1(X)}{48} c_1(F) \,.
\end{gather*}

As Chern classes take value in integral cohomology, it follows that, ignoring gravitational effects, in the presence of $G_2$ flux, $G_4/2\pi$ is generically half-integrally quantized, and $G_6/2\pi$ is quantized in multiples of $\frac{1}{6}$. Incorporating the $\hat{A}$-genus generically yields quantization in multiples of $\frac{1}{48}$ for both $G_4/2\pi$ and $G_6/2\pi$. In particular, for the cosets we are considering, the Pontrjagin classes are given by
\be
p\left(\frac{\mathrm{SU(3)}}{\mathrm{U(1)}\times \mathrm{U(1)}}\right)=1\quad, \quad p\left(\frac{\mathrm{Sp(2)}}{\mathrm{S(U(2)}\times\mathrm{U(1))}}\right)=(1+x^2)^4 \quad,\quad p\left(\frac{\mathrm{G}_2}{\mathrm{SU(3)}}\right)=1\;.
\ee
The first result follows from a theorem of Borel and Hirzebruch, according to which the Pontrjagin class of a coset $G/U$, with $U$ a maximal torus of $G$, is trivial. The latter two follow from the identification of the two cosets topologically with $\IC \IP^3$ and $S^6$ respectively. The $x$ that occurs is the generator of the integer cohomology of $\IC \IP^3$. It follows that $G_6/2\pi$ is quantized in multiples of $\frac{1}{6}$ for the cosets $\frac{\mathrm{SU(3)}}{\mathrm{U(1)}\times \mathrm{U(1)}}$ and $\frac{\mathrm{G}_2}{\mathrm{SU(3)}}$, and in multiples of $\frac{1}{12}$ for $\frac{\mathrm{Sp(2)}}{\mathrm{S(U(2)}\times\mathrm{U(1))}}$. For $\frac{\mathrm{G}_2}{\mathrm{SU(3)}}$, we can go further. In \cite{DMW}, the following mod 2 relation among Chern classes is derived
\ba
c_3(E) &=& c_1(E) c_2(E) + Sq^2 c_2(E) \hspace{0.2cm} \mod 2 \,.
\ea
Since $\frac{\mathrm{G}_2}{\mathrm{SU(3)}}$ has no 2- and 4-cohomology, it follows that $c_3(E)$ must be even for any vector bundle on this space (\cite{DMW} provide an index theory argument for this conclusion). We conclude that on this coset, $G_6$ is integrally quantized. These results are summarized in table~\ref{quantcond}.

\begin{table}
\begin{center}
\begin{tabular}{r|c|c|c}
           & $\frac{\mathrm{SU(3)}}{\mathrm{U(1)}\times \mathrm{U(1)}}$ & $\frac{\mathrm{Sp(2)}}{\mathrm{S(U(2)}\times\mathrm{U(1))}}$  &  $\frac{\mathrm{G}_2}{\mathrm{SU(3)}}$	 \\[2mm]
\hline
$G_0$ & $\IZ$ &  $\IZ$ & $\IZ$ \\ [1mm]
$G_2$ & $\IZ$ &  $\IZ$ & $-$ \\ [1mm]
$G_4$ & $\frac{1}{2} \IZ$ &  $\frac{1}{12}\IZ$ & $-$ \\ [2mm]
$G_6$ & $\frac{1}{6} \IZ$ &  $\frac{1}{12}\IZ$ & $\IZ$ \\
\end{tabular}
\caption{Quantization conditions on fluxes.} \label{quantcond}
\end{center}
\end{table}

We  turn to the second question raised above: given an element of $H^*(X,\IQ)$ satisfying the integrality conditions just discussed, when does it lie in the image of the map $ \sqrt{\hat{A}} \, \ch(\cdot)$, thus qualifying as a viable choice of flux? We will not provide a general answer, but address the following two subquestions which will be relevant in the next subsection.
\paragraph{Is it possible to have only $G_0$ and $G_6$ non-vanishing?} It is a theorem (see e.g. Thm. V.3.25 in \cite{Karoubi}) that the map (\ref{AH}) provides an isomorphism when the domain is extended to rational $K$-theory, $K(X) \otimes \IQ$. It follows that any class in $H^*(X, \IQ)$ lifts to a fractional $K$-theory class. Multiplying our choice of $G_0$ and $G_6$ with an appropriate integer hence always provides a viable choice of flux.
\paragraph{Given $G_2=0$, which $G_4$ are permissible?} Let us consider the class $$ x=\frac{[G_4]}{2\pi} + \frac{p_1(X)}{48} \rank(F)\,. $$ Among the geometries we consider, $x$ differs from $[G_4]/(2\pi)$ only for $\IC \IP^3$. $x$ is integrally quantized whenever $G_2$ vanishes. For the two cosets with non-trivial 2- and 4-cohomology, this is the only restriction on $x$, i.e. $x$ can take values in all of $H^4(X,\IZ)$. As pointed out in \cite{DMW}, this situation arises whenever the cohomology of the manifold is generated in second degree. If we call the generators $x_i$, line bundles $L_i$ exist with $c_1(L_i)=x_i$. The $K$-theory classes $x_{ij}=L_i \otimes L_j - L_i \oplus L_j$ can then be used as building blocks for lifting $x$ to a $K$-theory class, by
\ba
\ch(x_{ij}) &=& x_i x_j + \frac{1}{2}(x_i^2 x_j + x_i x_j^2) \,.
\ea
If we choose $G_0$-flux as a multiple of $12$, we can ignore the gravitational contribution which accounts for the difference between $x$ and $[G_4]/(2\pi)$ in the case of $\IC \IP^3$. Then, $[G_4]/(2\pi)$ takes values in $H^4(X,\IZ)$ also for this case.

\subsection{The solution}\label{TheSolution}
The $\cN=1$ supersymmetry conditions for an AdS$_4$ vacuum with internal SU(3) structure have been determined by \cite{LustTsimpis} (see \cite{GMPT2,GMPT3} for generalization to the SU(3)$\times$SU(3) structure context). A non-trivial warp factor is not allowed, and the dilaton $\phi$ has to be constant. Furthermore, in our conventions\footnote{Our supergravity field strengths are as in \cite{Democratics}. We derive the susy conditions starting from an ansatz for the two type IIA susy parameters $\epsilon^{1}, \epsilon^{2}$ which assigns negative chirality to $\epsilon^1$ and positive chirality to $\epsilon^2$. For the gamma matrices and the SU(3) structure we adopt the conventions listed in subsection$\:$A.2 of \cite{ReducingSU3SU3}. The resulting equations (\ref{eq:SU3susyConditions}) and the SU(3) torsion classes differ from the ones in \cite{KoerberLustTsimpis} by just a few minus signs. The factor of $(-1)^s=\pm 1$ arises in the following equations as unlike \cite{LustTsimpis}, we have fixed the phase of $\Omega$ once and for all in (\ref{eq:JandOmForCoset}); see also \cite{GMPT3}. Both signs are consistent with supersymmetry.} the equations governing the H-field and the internal RR field strengths read
\bea
\label{eq:SU3susyConditions} H &=& (-1)^s \frac{2m}{5} e^\phi \, \re \Omega \,, \label{ns}\\
\nnb F_0 &=& m \,, \hspace{1cm} F_2 = -\frac{f}{9}J + (-1)^s ie^{-\phi} W_2 \,,  \hspace{1cm}F_4 = \frac{3m}{10} J \wedge J \,, \hspace{1cm} F_6 = \frac{f}{6} J \wedge J \wedge J \,, \label{rr}
\eea
where the only non-vanishing purely imaginary torsion classes are $W_1 = (-1)^s \frac{4i}{9}e^\phi f$ and $W_2$. The only Bianchi identity which is not automatically satisfied is $dF_2 - HF_0=0$. This imposes 
\be
d W_2 = ie^{2\phi}\big(\frac{2}{27}f^2 - \frac{2}{5}m^2\big) \re \Omega \,. \label{bianchi}
\ee
The AdS cosmological constant is determined by
\be\label{eq:AdScurvature}
\Lambda = - 3 e^{2\phi} \left( \frac{m^2}{25}+\frac{f^2}{9}\right) \,.
\ee
Following work of \cite{TomasielloTwistor}, \cite{KoerberLustTsimpis} showed that these equations can be solved on the cosets we introduced in the previous section, by expanding all fields in forms invariant under the left group action. We will repeat this analysis, but parametrize the solutions by the fluxes $[G]$, as introduced in the previous subsection, rather than the parameter $f$ and the dilaton. This is the favored approach as it allows us to take flux quantization into account naturally (from a 4d point of view, the distinction between fluxes and parameters such as $f$ and the dilaton is most striking, as the former correspond to charges, the latter to VEVs; in 10d, while fluxes can also be considered as VEVs, they are distinguished by encoding topological information).

We will focus on $\frac{\textrm{SU(3)}}{\textrm{U(1)}\times \textrm{U(1)}}$ for concreteness. This example is the most rich among the three cosets we are considering, as it has the largest set of left-invariant forms, and the largest cohomology.

The ansatz (\ref{eq:JandOmForCoset}) already led to the expressions (\ref{eq:TorsionClassesExplicit}) for $W_1$ and $W_2$ in terms of the metric parameters $v^a$. It will prove convenient for this section to express the internal component $b$ of the B-field using the closed 2-forms (\ref{closed2}),
\ba
b &=& b'^1 \omega'_1 + b'^2 \omega'_2 + b'^3 \omega_3 \,.
\ea
Thus, $b'^1$ and $b'^2$ capture topological information about the $B$-field, while $b'^3$ enters in $H$. Likewise, our ansatz for $G$ is
\ba
G_0 &=& m \,, \\
G_2 &=& m'^{1} \omega'_1 + m'^{2} \omega'_2 \,, \\
G_4 &=&  -e_1 \tomega^1 - e_2 \tomega^2 - \txi \, d \beta \,, \\
G_6 &=& - e \tomega^0 \,. 
\ea
The equations of motion for $G$ are complicated, and are encoded in the equations (\ref{eq:SU3susyConditions}). By contrast, the Bianchi identities are already guaranteed by the ansatz (hence the use of primed forms).

To solve (\ref{eq:SU3susyConditions}) in terms of the flux parameters, we begin by solving (\ref{bianchi}) in term of $\phi$, invoking the relation between $W_1$ and $f$,
\ba
e^{2\phi} &=& \frac{5}{16 m^2 v^1 v^2 v^3} [6\sum_{a<b} v^a v^b- 5 \sum(v^a)^2]\,.
\ea
For the rest of this section, $\phi$ will denote this solution.

Utilizing the equation for $H$, this allows us to solve for $b'^3$ in terms of the metric parameters,
\ba
b'^3 &=& (-1)^{s+1} \frac{4m}{5} \sqrt{v^1 v^2 v^3} e^{\phi} \\
&=&   (-1)^{s+1} \frac{m}{|m|} \sqrt{5 \left(6\sum_{a<b} v^a v^b- 5 \sum(v^a)^2 \right)}  \,.
\ea

We next want to solve for $f$, starting with
\ban
F_6 &=& G_6 + B \wedge G_4  +\frac{1}{2} B^2 \wedge G_2 + \frac{1}{3!}B^3 \wedge G_0 = \frac{f}{6} J \wedge J \wedge J \,.  \label{f6}
\ean
We eliminate the $B^3$ term via
\ba
&&F_4 = G_4 + B\wedge G_2 + \frac{1}{2} B \wedge B \wedge G_0 = \frac{3m}{10} J \wedge J \\
& \Leftrightarrow& mB^3 = \frac{3m}{5} B \wedge J \wedge J - 2 B \wedge G_4 - 2B^2 \wedge G_2 \,.
\ea
Hence,
\ba
\frac{f}{6} J \wedge J \wedge J &=& G_6 + \frac{2}{3} B \wedge G_4 + \frac{1}{6} B^2 \wedge G_2 + \frac{m}{10} B\wedge J \wedge J \,,
\ea
and substituting $f$ into 
\ba
F_2 = G_2 + B\wedge G_0 = -\frac{f}{9}J + (-1)^{s} ie^{-\phi} W_2^-
\ea
yields three equations which can be solved for $b'^1$, $b'^2$ and $\txi$,
\ba
b'^1 &=& (-1)^s \frac{(5 v^1 - 3 (v^2 + v^3)) \sqrt{v^1 v^2 v^3}}{4 v^2 v^3 m}e^{-\phi} - \frac{m'^1}{m} \,,\\
b'^2 &=& (-1)^s \frac{(5 v^2 - 3 (v^1 + v^3)) \sqrt{v^1 v^2 v^3}}{4 v^1 v^3 m}e^{-\phi} - \frac{m'^2}{m}
\ea
We omit the expression for $\txi$, which is lengthy and not illuminating.

At this stage, we have expressed $\txi, b'^a, e^\phi$ in terms of $v^a$. Substituting these into the $F_4$ equation,
\ban
G_4 + G_2 \wedge B + \frac{1}{2} B \wedge B \wedge G_0 = \frac{3m}{10} J \wedge J \,, \label{g4eq}
\ean
yields three independent equations for $v^a$, two of which take the simple form
\ban
 \frac{(v^1-v^3)(v^1 v^2+ v^2 v^3- 3 v^1 v^3)}{v^1 v^3}-e^{2\phi} (\frac{m e_1}{I} +m'^2(2 m'^1+m'^2)) &=& 0 \,,  \nn\\
\frac{(v^2-v^3)(v^1 v^2+ v^1 v^3- 3 v^2 v^3)}{v^2 v^3}-e^{2\phi} (\frac{m e_2}{I} +m'^1(m'^1+2m'^2)) &=& 0 \,. \label{g4imp}
\ean
The main new feature we wish to demonstrate, as compared to the Nearly K\"ahler analysis of \cite{KashaniPoorNearlyKahler}, is the presence of several supersymmetric vacua of a given theory, i.e. upon a fixed choice of fluxes. This phenomenon already occurs at $e_a = m'^a = 0$, which is a permissible choice of flux by the previous subsection. The third equation following from (\ref{g4eq}) here takes the form
\ba
15 e^\phi\sqrt{v^1 v^2 v^3} \, e +(-1)^{s+1} 8 I\, v^2 v^3 (v^2 v^3 -3  v^1 v^2- 3 v^1 v^3) = 0 \,.
\ea
It is easy to see that this system of equations has, aside from the Nearly K\"ahler \hbox{solution at}\footnote{Note that physicality (positivity of $v^a$) determines the appropriate choice of $s$ depending on the sign of $e$.} $$v^1 = v^2 = v^3 = \frac{\sqrt{15}}{2}\left(\frac{1}{20 I} \left |\frac{e}{m}\right |  \right)^{\frac{1}{3}}\,, $$ the solution $$v^1 = v^2 = 2 v^3 = \frac{\sqrt{15}}{4}\left(\frac{1}{2 I} \left |\frac{e}{m}\right| \right)^{\frac{1}{3}}\,,$$ as well as two others which arise upon cyclic permutation of $v^1, v^2, v^3$.

The symmetry between the three metric parameters $v^1, v^2, v^3$ can be broken by considering backgrounds with $G_4$ flux. E.g., maintaining $G_2 =0$, we obtain from (\ref{g4imp})
\ba
e_1 \neq 0 &\rightarrow& v^1 \neq v^3 \,,\\
e_2 \neq 0 &\rightarrow& v^2 \neq v^3 \,,\\
e_1 \neq e_2 &\rightarrow& v^1 \neq v^2 \,.
\ea
We have checked numerically that e.g. at $e_1 \neq 0, e_2 =0$, solutions with $v^2 = v^3$ exist.

\section{The dimensional reduction}  \label{consistent}

\subsection{The truncation scheme}  \label{consistentint}

As announced, we will adopt a reduction prescription in which the higher dimensional supergravity fields are expanded on a basis for the left-invariant tensors admitted by the coset. This expansion basis was introduced in subsect. \ref{ExpBasis} for the three cosets (\ref{eq:OurCosets}).

We stress again that this $G$-invariant truncation does not coincide with a massless Kaluza-Klein ansatz. We can illustrate the differences between the two schemes e.g. by considering the gauge vectors of the dimensionally reduced theory arising from the decomposition of the higher dimensional metric. The conventional massless Kaluza-Klein ansatz associates a gauge vector of the truncated theory to each Killing vector on the compact manifold, the gauge symmetry being inherited from the reparameterization invariance of the higher dimensional spacetime.\footnote{In principle, non-vanishing background values of the non-metric supergravity fields may break the gauge symmetry to a subgroup of the isometry group, however this is guaranteed not to happen as far as these vevs are invariant under the isometries \cite[pag.$\:$16]{DuffNilssonPopeKKreview}.} On the other hand, the $G$-invariant ansatz preserves just a subgroup of the full isometry group of the internal manifold $G/H$. The theory of compact left coset spaces endowed with a left-invariant metric (such are the cosets we consider) states that in general the isometry group of $G/H$ is $G\times N(H)/H$, where $N(H)$ is the normalizer of $H$ in $G$, defined as $N(H):=\{g\in G : gH= Hg \}\,$. The $G$ factor in $G\times N(H)/H$ is associated with the left action of $G$ on the coset, while the $N(H)/H$ factor derives from the right action of $G$. The Killing vectors generating the right isometries are left-invariant, while this is not the case for the ones generating the left isometries.\footnote{A detailed discussion of the isometries of $G/H$ can be found in section$\:$2 of ref.$\:$\cite{CastellaniRomansWarner}.} It follows that a left-invariant reduction ansatz keeps only the former, and the gauge group descending from the higher dimensional metric sector is just $N(H)/H$.

For the cosets we consider the $G$-invariant ansatz is particularly simple, because $N(H)/H$ turns out to be trivial. This can be seen either by observing that $\mathrm{rank}\, G = \mathrm{rank}\, H$ \cite{MuellerStuckl}, or by noticing that our cosets do not admit left-invariant vectors at all. We conclude that no gauge vectors will descend from the dimensional reduction of the type II supergravity NSNS sector, and the whole (abelian) gauge group will be provided by the RR sector. This is analogous to what is realized in Calabi-Yau compactifications.

Though physically well motivated, dimensional reductions based on the full massless KK ansatz have a drawback: they are generically inconsistent \cite{DuffNilssonPopeWarner,DuffNilssonPopeKKreview}. Rare exceptions are known, an example being the $S^7$ reduction of \cite{DeWitNicolaiConsistency} (see \cite{ConsistentSphere} for a discussion of consistent KK sphere reductions). The $G$-invariant reduction scheme is instead believed to provide consistent truncations, due to the fact that the preserved invariant fields never generate the truncated non-invariant modes. A further argument for consistency is that the substitution of a $G$-invariant ansatz guarantees the dropping of the dependence on the internal coordinates $y$ from the higher dimensional Lagrangian, see e.g. \cite{DuffPopeConsistentKK,DuffNilssonPopeKKreview} for more details. The consistency of the $G$-invariant scheme was explicitly shown in ref.$\:$\cite{CoquereauxJadczykConsistency} for a reduction of the pure gravity action. Recent related discussions can be found in \cite{Chatzistavrakidis2007} (for coset space reductions of Einstein-Yang-Mills theories), in \cite{HullReidEdwardsStringTwistedTori, Dall'AgataPrezas-ScherkSchwarz} (for Scherk-Schwarz reductions on group manifolds), and in \cite{GauntlettVarela, GauntlettKimVarelaWaldram} (for consistent reductions on spaces supporting AdS solutions, and their relation with the dual SCFT). However, an explicit check of consistency in the context of SU(3) structure compactifications with fluxes had not been performed to date. In subsection$\:$\ref{Consistency} we will work out the reduction of the higher dimensional equations of motion in detail, and prove the consistency of the truncation of the full type IIA bosonic sector for the cosets (\ref{eq:OurCosets}).

\subsection{The 4d action}\label{4daction}

Following the reduction prescription for type IIA on SU(3) structure manifolds initiated in \cite{generalized mirror symmetry}, the complete 4d gauged $\cN=2$ bosonic action has by now been derived \cite{GaugingHeisenberg, TomasMirrorSymFl, GLW1, HousePalti, KashaniPoorMinasian}. Here, we will use the notation of ref.$\:$\cite{ReducingSU3SU3}. Separating the contributions of the NSNS and RR sectors, the action $S^{(4)}$ arising from a reduction on our cosets (\ref{eq:OurCosets}) reads $S^{(4)}=S^{(4)}_{\mathrm{NS}} + S^{(4)}_{\mathrm{RR}}$, with
\bea\label{eq:S4NS}
S^{(4)}_{\mathrm{NS}} 
\!\!&=&\!\!\int_{M_4}  \Big(\,\frac{1}{2} R_{4}*1  -\frac{1}{4}e^{-4\varphi}dB\wedge *dB -  d\varphi \wedge *d\varphi - \cl G_{ab} dt^a \wedge * d\bar t^b - V_{\mathrm{NS}}*1\,\Big),\\ [5mm]
\nnb S^{(4)}_{\mathrm{RR}}\!\! &=&\!\! \int_{M_4} \Big\{\;\frac{1}{4}\im \cl N_{AB} F^A \wedge * F^B + \frac{1}{4}\re \cl N_{AB} F^A \wedge F^B - \frac{e^{2\varphi}}{4} (D\xi \wedge * D\xi + d\tilde\xi \wedge * d\tilde\xi)\\ [1mm]
\label{eq:S4RR} &&\;\; +\; \frac{1}{4}dB\wedge \big[ \xi d\tilde\xi - \tilde\xi D\xi + 2 e_{A}A^A +  \tilde \xi \,q_a A^a \,\big] - \frac{1}{4} m^A e_{A} B\wedge B -  V_{\mathrm{RR}} *1\,\Big\} \,.
\eea
The different quantities appearing in this 4d action are introduced in appendix$\:$\ref{DetailsDimRed}, where we also give some details about the derivation from the higher dimensional supergravity. The 4d degrees of freedom descending from the NSNS sector are the metric $g_{\mu\nu}$, the 2--form $B$, the complex scalars $t^a=b^a+iv^a$ and the 4d dilaton $\varphi$, defined in (\ref{eq:Def4dDilaton}). The RR sector yields the scalars $\xi$ and $\tilde \xi$ introduced in the first line of (\ref{eq:4dDOF}), as well as the gauge potentials $A^A$, whose modified field strengths $F^A$ are defined in (\ref{eq:ModifFieldStrp=0=qCase}) (recall that the index $A$ runs over $(0,a)\,$). 

The $\cN=2$ action $S^{(4)}$ contains the gravitational multiplet $(g_{\mu\nu},A^0)$, a number of vector multiplets $(t^a,A^a)$ (see table \ref{DetailsCosets} for the coset dependent range of $a$), and one tensor multiplet $(B,\varphi,\xi,\tilde\xi)$. When $m^A=0$ the antisymmetric tensor $B$ becomes massless and can be dualized to a scalar, yielding the universal hypermultiplet. From $D\xi= d\xi -  q_a A^a$ it follows that $\xi$ is charged under the $A^a$, the charges being provided by the geometric fluxes $q_a$ given in table \ref{DetailsCosets}. The graviphoton $A^0$ instead does not participate to this gauging (due to the fact that the compactification manifolds (\ref{eq:OurCosets}) do not allow for a flux of the NSNS 3--form \cite{generalized mirror symmetry}).

The special K\"ahler metric $\cl G_{ab}$ governing the kinetic terms for the scalars in the vector multiplets is given in table \ref{DetailsCosets}, and further discussed in subsection \ref{SpKahlerGeo} of the appendix, together with the period matrix $\cl N_{AB}$ describing the kinetic and topological terms for the gauge potentials. 

The full 4d scalar potential reads $ V =  V_{\mathrm{NS}} +  V_{\mathrm{RR}}$. Reduction of the internal NSNS sector on our coset spaces yields\footnote{For any pair of forms $ P, Q$ of degree $k$ we define the contraction $P\,\lrcorner\,  Q:= \frac{1}{k!}P_{m_1\ldots m_k}Q^{m_1\ldots m_k}$. In our conventions for the Hodge $*$, we have $\big( P\,\lrcorner\,  Q\big) *1 = P\wedge  * Q$. This also holds for the 10d spacetime equations of the forthcoming subsection.}
\bea
\nnb V_{\mathrm{NS}} &\equiv&  -\frac{e^{2\varphi}}{2} \big( R_6 -\frac{1}{2}H \lrcorner H\big)\\ [1mm] 
\label{eq:VNS} &=&  \frac{e^{2\varphi}}{4 V\!ol}q_aq_b\big(\,\cl G^{ab} -3v^av^b  + b^ab^b \,\big)\;,
\eea
where the 6d Ricci scalar $R_6$ has been evaluated in terms of the torsion classes expressed in eq. (\ref{eq:TorsionClassesExplicit}) via the formula\footnote{An equivalent expression for $R_6$ was given in \cite{MuellerStuckl} using a general formula relating the Riemann tensor of $G/H$ to the structure constants of $G$. The 4 factor mismatch we have with respect to that expression is due to the different normalization of the SU(3) structure constants already mentioned in footnote$\:$\ref{ftn:FtnStrCnst}.} \cite{BedulliVezzoni}
\be\label{eq:RfromTorsion}
R_6= \frac{15}{2}|W_1|^2 - \frac{1}{2} W_2\lrcorner \overline W_2\;,
\ee
while for the internal NSNS 3--form we have $ H = d_6b = b^aq_a\alpha$.

The RR contribution to the scalar potential, obtained from the general expression given in eq.$\:$(\ref{eq:V_RRgeneral}) of the appendix, is
\be\label{eq:DefV_RR} 
V_{\mathrm{RR}}\!= -\frac{e^{4\varphi}}{ 4}\big[ m^A\im\mathcal{N}_{AB}m^B + (e_A+q_A\tilde\xi - m^C\re\mathcal{N}_{CA})(\im\mathcal{N})^{-1\,AB}(e_B+q_B\tilde\xi - \re\mathcal{N}_{BD}m^D) \big],
\ee
where $q_A=(0,q_a)$. Notice that while $\tilde \xi$ appears in the potential, the other RR scalar $\xi$ is a flat direction (however, $\xi$ is not a modulus, since it is charged under the $A^a$). Since the matrix $\im\mathcal{N}$ is negative, $V_{\mathrm{RR}}$ is positive semi-definite.

\subsection{Consistency of the truncation}\label{Consistency}
We now prove the consistency of the dimensional reduction leading to the 4d action $S^{(4)}$ introduced in the previous subsection. To this end, we plug the $G$-invariant reduction ansatz into the bosonic equations of motion (EoM) of type IIA supergravity, and show that these yield the EoM following from the reduced action $S^{(4)}$. 

The reduction of the equations for the RR degrees of freedom was already described in the general analysis of \cite{ReducingSU3SU3} and is summarized, for the specific compactification on the coset spaces (\ref{eq:OurCosets}), in subsection$\:$\ref{ReductionRRsector} of the appendix. In fact, the piece (\ref{eq:S4RR}) of the 4d action has been established requiring its compatibility with the EoM for the 4d fields $A^A, \xi,\tilde \xi$ as obtained from the higher dimensional equations (\ref{eq:Bianchi10dG}), (\ref{eq:SelfDuality10dG}). It follows that, as far the RR sector is concerned, the reduction is consistent by construction.

Hence, we just have to analyse the equations of motion for the NSNS degrees of freedom, namely the $B$-field, the Einstein and the dilaton equations. For the democratic formulation of type IIA supergravity \cite{Democratics} in string frame, these read
\be\label{eq:10dBEoM}
d(e^{-2\phi} *\hat H) -\frac{1}{2}[{\bf\hat F}\wedge *{\bf\hat F}]_8 \;=\; 0\;,
\ee
\be\label{eq:10dEinstein}
\hat R_{MN}  + 2 \hat\nabla_M\partial_N \phi -\frac{1}{2}\iota_M\hat H\lrcorner \iota_N\hat H - \frac{e^{2\phi}}{4}\sum_{k=0,2}^{10} \iota_M\hat F_{(k)}\lrcorner \iota_N\hat F_{(k)}\;=\; 0\;,
\ee
\be\label{eq:10dDilatonEoM}
\hat R - \frac{1}{2}\hat H\lrcorner\hat H + 4\big(\hat\nabla^2 \phi - \partial_M\phi \hat\partial^M\phi \big) = 0\;,
\ee
where the hat denotes 10d quantities, ${\bf \hat F}\equiv\sum_{k=0,2}^{10}\hat F_{(k)}$ is the sum of the RR field-strengths, and $M,N$ are 10d spacetime indices.

\vskip .5cm

\ul{\it $\hat B$-field EoM}\vskip .2cm

The $\hat B$-field EoM (\ref{eq:10dBEoM}) is an 8--form equation. Its expansion in the left-invariant forms on $M_6$ yields two independent equations: the first exhibiting two indices along 4d spacetime $M_4$ and 6 indices along $M_6$, and the second with 4 indices along $M_4$ and 4 indices along $M_6$. We get no equation with 5 indices along $M_6$ due to the absence of invariant 5-forms on the cosets (\ref{eq:OurCosets}). Concretely, recalling (\ref{eq:SelfDuality10dG}) we rewrite the RR piece of (\ref{eq:10dBEoM}) as
\be\nnb
[{\bf\hat F}\wedge *{\bf\hat F}]_8 \;=\; [{\bf\hat F}\wedge \lambda({\bf\hat F})]_8 \;=\; [{\bf\hat G}\wedge \lambda({\bf\hat G})]_8\;.
\ee
Expanding $\hat B$ as in (\ref{eq:SplitB}) and ${\bf\hat G}$ as in (\ref{eq:ExpansionWithG}), we see that eq. (\ref{eq:10dBEoM}) reduces to
\be
\label{eq:4dBEoM} \Big[\, d(e^{-4\varphi}*dB)\, +\, G_{(0)}^A \tilde G_{(2)A}  - \tilde G_{(0)A} G_{(2)}^A +  \tilde G_{(1)}\wedge G_{(1)} \Big] \tilde\om^0 = 0
\ee
and
\bea
\label{eq:EqForTheb} &-&\!\! 4 d_4(\cl G_{ab}*_{4} d_4 b^b)\tilde\om^a \:+\: e^{-2\phi + 4\varphi} vol_4\wedge d_6(*_6d_6 b)\:+ \\ [1mm]
\nnb &+&\!\! \Big[ G_{(0)}^0 \tilde G_{(4)a} + G_{(4)}^0\tilde G_{(0)a}  -\cl K_{abc} G_{(0)}^b G_{(4)}^c - G_{(2)}^0 \wedge\tilde G_{(2)a}  + \frac{1}{2}\cl K_{abc} G_{(2)}^b \wedge G_{(2)}^c \Big]\tilde \om^a =0\;,
\eea
where the 4d forms $G_{(p)},\tilde G_{(p)}$ are expressed in (\ref{eq:4dDOF}), and we used $\om_a\wedge \om_b=-\cl K_{abc}\tilde\om^c$, with the $\cl K_{abc}$ given in (\ref{eq:IntNumb}).

Eq.$\:$(\ref{eq:4dBEoM}) provides the EoM for the 2--form $B$ in 4d. It already appeared in section$\:$5 of ref.$\:$\cite{ReducingSU3SU3}, where it was employed in order to deduce the 4d action $S^{(4)}$ written in subsection \ref{4daction} above. It follows that, on the same footing as the RR equations, consistency of this equation with the action $S^{(4)}$ is guaranteed by construction. 

Eq.$\:$(\ref{eq:EqForTheb}) (which was not analysed in \cite{ReducingSU3SU3}) corresponds to the EoM for the 4d scalars $b^a$ defined by the expansion of the internal B-field $b$ on the basis 2--forms. Using $d_6*_6 d_6 b =q_bb^bq_a\tilde\om^a $, substituting the expressions (\ref{eq:4dDOF}) for $G_{(2)},\tilde G_{(2)},G_{(4)},\tilde G_{(4)}$ and the definition (\ref{eq:ModifFieldStrp=0=qCase}) of $F^A$, eq.$\:$(\ref{eq:EqForTheb}) reads
\bea
\nnb && 4 \nabla_\mu(\cl G_{ab}\partial^\mu b^b) \;-\; e^{2\varphi}\frac{q_bb^bq_a}{  V\!ol} \;-\; \im\cl N_{aB} * (F^0\wedge *F^B) - \re \cl N_{aB}* (F^0\wedge F^B )\\ [1mm]
\nnb &+&\!\!\! \frac{1}{2} \cl K_{abc}*(F^b\wedge F^c)  + e^{4\varphi}\big[ G_{(0)}^0 ( \im\cl N G_{(0)} - \re\cl N L )_a -\tilde G_{(0)a} L^0 + \cl K_{abc} G_{(0)}^b L^c \big]\,=\,0,
\eea
where we denote $L \equiv (\im\cl N)^{-1}(\tilde G_{(0)}-\re\cl N G_{(0)})$. Recalling the form of $\im \cl N$ and $\re \cl N$ in (\ref{eq:ImN}) and (\ref{eq:ReN}), as well as $V_{\mathrm{NS}}$ in (\ref{eq:VNS}) and $V_{\mathrm{RR}}$ in (\ref{eq:V_RRgeneral}), one checks that this equation can be reformulated as
\be
\nnb 2\nabla_\mu(\cl G_{ab}\partial^\mu b^b) -\frac{1}{4}\partial_{b^a}\im\cl N_{BC} *( F^B \wedge *F^C )  -\frac{1}{4}\partial_{b^a}\re\cl N_{BC} *( F^B \wedge F^C )  - \partial_{b^a}(V_{\mathrm{NS}} + V_{\mathrm{RR}} )  = 0
\ee
which is precisely the EoM obtained varying $S^{(4)}$ in (\ref{eq:S4NS}), (\ref{eq:S4RR}) with respect to $b^a$.

\vskip .5cm

\ul{\it 10d Einstein equation}\vskip .2cm

We first deal with the term $\hat R_{MN}  + 2 \hat\nabla_M\partial_N \phi$ in eq.$\:$(\ref{eq:10dEinstein}). Starting from the $G$--invariant metric ansatz (\ref{eq:10dMetricAnsatzCoset}) and recalling that the 4d dilaton $\varphi(x)$ satisfies (\ref{eq:DerivativeVarphi}), we derive the following decomposition\footnote{The non-vanishing higher dimensional Christoffel symbols are:
\be\nnb
\hat \Gamma^\rho_{\mu\nu} = \Gamma^\rho_{\mu\nu} + \partial_\mu\varphi \delta_\nu^\rho + \partial_\nu \varphi\delta_\mu^\rho - g_{\mu\nu}\partial^\rho\varphi\;\;,\quad
\hat \Gamma^\rho_{mn} = -\frac{1}{2}e^{-2\varphi}\partial^\rho g_{mn}\;\;,\quad  \hat \Gamma^p_{\mu n} = \frac{1}{2}g^{pq}\partial_\mu g_{nq}\;\;,\quad \hat \Gamma^p_{mn} = \Gamma^p_{mn}\;.
\ee
In the derivation of $\hat R_{\mu  n}=0$ we assume $\nabla_m e^{\ul p}_{\;\,n}=0$.}
\bea
\nnb \hat R_{\mu\nu} + 2\hat\nabla_\mu\partial_\nu\phi &=& R_{\mu\nu} - \frac{1}{4} g^{\ul{mp}}g^{\ul{nq}} \partial_\mu g_{\ul{mn}}\partial_\nu g_{\ul{pq}} - 2\partial_\mu\varphi \partial_\nu\varphi   - g_{\mu\nu}\nabla_{\!4}^2\varphi  \;, \\ [2mm]
\nnb \hat R_{\mu  n} &=& 0\; =\;\hat\nabla_\mu\partial_n \phi\;,\\
\label{eq:RicciTensorSimplified}\hat R_{\ul{mn}} + 2 \hat\nabla_{\ul m}\partial_{\ul n}\phi &=&   R_{\ul{mn}} + \frac{1}{2} e^{-2\varphi}\big( g^{\ul{pq}} \partial_\mu g_{\ul{mp}}\partial^\mu g_{\ul{nq}}  - \nabla_{\!4}^2 g_{\ul{mn}}\big)\;.
\eea
Taking the trace, we get
\be\label{eq:SimplifiedRicciScalarRed} \hat R  +4 \hat\nabla^2\phi   - 4\partial_M\phi \hat\partial^M\phi \; =\; e^{-2\varphi}\big(\, R_4 + e^{2\varphi} R_6 - \frac{1}{4} g^{\ul{mp}}g^{\ul{nq}}\partial_\mu g_{\ul m\ul n}\partial^\mu g_{\ul p\ul q} -  2\nabla_{\!4}^2\varphi - 2\partial_\mu\varphi \partial^\mu\varphi\big)\,.
\ee
In the previous expressions, quantities labeled with $4$ or $6$ are associated to $(M_4, g_{\mu\nu})$ or $(M_{6}, g_{mn})$ respectively. The 4d indices on the r.h.s. are raised using the rescaled metric $g^{\mu\nu}$ of eq.  (\ref{eq:10dMetricAnsatzCoset}). Notice that all the terms depend just on $x^\mu$: indeed, thanks to $G$-invariance, the whole dependence on the internal coordinates drops out.

Let us now consider the $\mu \nu$ components of the 10d Einstein equation (\ref{eq:10dEinstein}). Using (\ref{eq:RicciTensorSimplified}), (\ref{eq:SimplifiedRicciScalarRed}) we find (we reinstate in the Einstein equation the term proportional to $\hat g_{\mu\nu}$, which actually vanishes thanks to the dilaton EoM (\ref{eq:10dDilatonEoM})$\,$),
\bea\label{eq:10dEinsteinMuNu1}
\nnb \hat R_{\mu\nu}  \!\!&+&\!\! 2 \hat\nabla_\mu\partial_\nu \phi - \frac{1}{2}\iota_\mu\hat H\lrcorner \iota_\nu\hat H - \frac{1}{2}\hat g_{\mu\nu}\Big(\hat R + 4\hat\nabla^2\phi -4\partial_\rho\phi \hat\partial^\rho\phi - \frac{1}{2}\hat H^2 \Big) \;=\\ [1mm]
\label{eq:PieceOfEinstWithMuNuIndices} &=& R_{\mu\nu} - \frac{1}{4} e^{-4\varphi} H_{\mu\rho\sigma}H_{\nu}^{\phantom{\nu}\rho\sigma} - 2\partial_\mu \varphi\partial_\nu \varphi -2\cl G_{ab} \partial_{(\mu} t^a \partial_{\nu)} \bar t^b  \\ [1mm]
\nnb &-& \!  g_{\mu\nu} \Big( \,\frac{1}{2} R_{4} -\frac{1}{24}e^{-4\varphi}H_{\mu\nu\rho}H^{\mu\nu\rho} -  \partial_\mu\varphi \partial^\mu\varphi - \cl G_{ab} \partial_\mu t^a \partial^\mu \bar t^b  - V_{\mathrm{NS}}\, \Big)\;.
\eea
For the RR piece, taking into account all the terms of the expansion described in subsection \ref{ReductionRRsector} of the appendix, we arrive at
\bea\label{eq:10dEinsteinMuNu2} 
\nnb - \frac{e^{2\phi}}{4}\sum_{k=0}^{10} \iota_\mu\hat F_{(k)}\lrcorner \iota_\nu\hat F_{(k)} &=&  \frac{1}{2} \im\cl N_{AB} \iota_\mu F^A\lrcorner \iota_\nu F^B -  \frac{1}{2} e^{2\varphi} (D_\mu\xi D_\nu\xi + \partial_\mu\tilde \xi \partial_\nu\tilde\xi)\\
- g_{\mu	\nu} \Big\{\,\frac{1}{4}\!\!\!\!\!\!\!& &\!\!\!\!\!\!  \im\cl N_{AB} F^A\lrcorner F^B - \frac{e^{2\varphi}}{4}[ (D_\mu\xi)^2  + (\partial_\mu\tilde \xi\,)^2] - V_{\mathrm{RR}} \Big\}\;.
\eea
From (\ref{eq:10dEinsteinMuNu1}), (\ref{eq:10dEinsteinMuNu2}) we see that the equation arising from the $\mu \nu$ components of (\ref{eq:10dEinstein}) precisely reproduces the 4d Einstein equation following from $S^{(4)}$.

Since there are no left-invariant 1--forms on the cosets (\ref{eq:OurCosets}), the 10d Einstein equation with $\mu n$ indices is trivialized by our left-invariant truncation prescription, and does not yield any constraint at the 4d level. Indeed, one can check that all the $\mu n$ terms in (\ref{eq:10dEinstein}) vanish once the truncation ansatz is plugged in.

Finally, we study the purely internal components of (\ref{eq:10dEinstein}) in flat $\ul{mn}$ indices. Depending on which of the cosets (\ref{eq:OurCosets}) we consider, these yield just one, two or three 4d scalar equations, labeled by the index $a$. On our cosets, any left-invariant symmetric rank-2 tensor has the same diagonal structure as the invariant metric $g_{\ul{mn}}$ given in subsection \ref{ExpBasis}. Furthermore, the left-invariant Ricci tensor on coset spaces satisfies
$R_{\ul{mn}} \,=\, \frac{\partial \,}{\partial g^{\ul{mn}}}R_6 $. Focusing for definiteness on $\frac{\textrm{SU(3)}}{\textrm{U(1)}\times \textrm{U(1)}}$, we have (recall $\cl G^{ab}$ in table \ref{DetailsCosets})\vskip -2mm
\be\nnb R_{\ul{2a-1}\,\ul{2a-1}} \equiv R_{\ul{2a}\,\ul{2a}} \;=\; -\frac{1}{8}\cl G^{ab} \partial_{v^b}R_6\;,\qquad a=1,2,3\,.
\ee
Then, using the last line of (\ref{eq:RicciTensorSimplified}), we get
\be\label{eq:RedEinstmnPart1}
\hat R_{\ul{2a}\,\ul{2a}} + 2 \hat\nabla_{\ul{2a}}\partial_{\ul{2a}}\phi -\frac{1}{2} \iota_{\ul{2a}}\hat H\lrcorner \iota_{\ul{2a}}\hat H \,=\,  \frac{e^{-2\varphi}\cl G^{ab}}{4}\big[-2 \nabla_\mu(\cl G_{bc}\partial^\mu v^c) + \partial_{v^b}\cl G_{cd} \partial_\mu t^c\partial^\mu \bar t^d   + \partial_{v^b} V_{\mathrm{NS}} \big].
\ee
Concerning the RR term, a tedious computation gives
\be\label{eq:RedEinstmnPart2}
-\frac{e^{2\phi}}{4} \sum_{k=0}^{10} \iota_{\ul{2a}}\hat F_{(k)}\lrcorner \iota_{\ul{2a}}\hat F_{(k)}  \;=\;  \frac{e^{-2\varphi}\cl G^{ab}}{4}\big[\partial_{v^b}V_{\mathrm{RR}} -\frac{1}{4}\partial_{v^b}(\im\cl N_{CD}) F^C\lrcorner F^D \big]\;.
\ee
Analogous steps can be repeated for the cosets $\frac{\mathrm{Sp(2)}}{\mathrm{S(U(2)}\times\mathrm{U(1))}}$ and $\frac{\textrm{G}_2}{\textrm{SU(3)} }$, leading to the same r.h.s. of the equations here above.

From (\ref{eq:RedEinstmnPart1}), (\ref{eq:RedEinstmnPart2}) we conclude that the components of the 10d Einstein equation (\ref{eq:10dEinstein}) with two internal indices precisely match the \hbox{EoM for the scalars $v^a$ following from $S^{(4)}$:}
\bea
\nnb &-&\!\!2 \nabla_\mu(\cl G_{ab}\partial^\mu v^b) + \partial_{v^a}\cl G_{bc} \partial_\mu t^b\partial^\mu \bar t^c   + \partial_{v^a} (V_{\mathrm{NS}}+ V_{\mathrm{RR}}) -\frac{1}{4}\partial_{v^a}(\im\cl N_{BC}) F^B\lrcorner F^C\;=\; 0\;.
\eea
\vskip .2cm
\ul{\it Dilaton equation}\vskip .2cm

Subtracting the trace over the $\mu\nu$ components of (\ref{eq:10dEinstein}) from the 10d dilaton equation (\ref{eq:10dDilatonEoM}), we eventually obtain
\be\label{eq:ReducedDilatonEq}
2\nabla_4^2\varphi + \frac{1}{6}e^{-4\varphi}H_{\mu\nu\rho}H^{\mu\nu\rho} - \frac{e^{2\varphi}}{2} \big[\,(D_\mu\xi)^2 + (\partial_\mu\tilde\xi)^2\,\big] - 2V_{\mathrm{NS}} - 4V_{\mathrm{RR}} = 0\;,
\ee
which is the EoM for the 4d dilaton $\varphi$ following from $S^{(4)}$. 

This concludes the consistency proof of the dimensional reduction.

\section{The 4d potential via $\cN=2$} \label{4dpotential}
In this section, we recast the scalar potential obtained in (\ref{eq:VNS}) and (\ref{eq:DefV_RR}) in 4d $\cN=2$ language. In this framework, given the prepotential $\cl F$ governing the special geometry data of the vector multiplet sector and the quaternionic metric $h_{uv}$ of the hypermultiplet sector, the potential is uniquely determined by the gauged isometries of $h_{uv}$. This structure allows us to incorporate string loops into our considerations, which correct the hypermultiplet metric. As the 4-dimensional quaternionic metrics with the isometry structure imposed by our compactifications are highly constrained, we use the results of \cite{Calderbank, AMTV} to write down the general form of the all-loop string corrected potential in subsection \ref{allloop}. We analyse this potential further in subsection \ref{desitvac}.

The general form of the potential in 4d $\cN=2$ gauged supergravity is \cite{N=2review,Michelson,N=2withTensor1,N=2withTensor2}
\ban
V &=& 4 e^K h_{uv} ( X^A k_A^u - \tilde{k}^{uA} \cl F_A) ( \bar{X}^B k_B^u - \tilde{k}^{uB} \bar{\cl F_B}) \nn \\
& & -\left[ \frac{1}{2} (\im \cN)^{-1 \,AB} + 4 e^K X^A \bar{X}^B \right] ( \cP_A^x - \tP^{xC} \cN_{CA}) ( \cP_B^x - \tP^{xD} \bar{\cN}_{DB}) \,. \label{potential}
\ean
The coordinates $X$, the prepotential $\cl F$, and the gauge coupling matrix $\cl N$ encode special geometry data and are discussed further in appendix \ref{DetailsDimRed}.
$h_{uv}$ refers to the universal hypermultiplet metric, which is expressed in terms of the quaternionic vielbein components as 
\ba
h &=& u \otimes \bar{u} + v \otimes \bar{v} \,.
\ea
We will denote the quaternionic coordinates collectively by $q^u$. $k_A^u$ and $\tilde{k}^{uA}$ are the components of the Killing vectors describing the isometries of the hypermultiplet metric being gauged by the $A^{\rm th}$ gauge vector. The Sp(1) factor $\omega$ of the spin connection of the hypermultiplet metric enters in the potential via its relation to the Killing prepotentials. For the case that the 3 components of the curvature of $\omega$ each are invariant under an isometry $k^u \partial_{q^u}$ of the metric, the corresponding Killing prepotential is given by \cite{Michelson, D'AuriaFerrFre}
\be\label{eq:RelPomega}
\cP^x\, =\, \omega^x_u k^u \,.
\ee
In this case, one can rewrite the potential in a more convenient form. Introducing
\ba
Q_A^u = k_A^u - \tk^{uB} \cN_{BA} \,,
\ea
we obtain
\be
V= Q_A^u \bar{Q}_B^v \Big[ 4 e^K X^A \bar{X}^B \big( u \otimes \bar{u} + v \otimes \bar{v} \big)_{uv}   -\big(4 e^K X^A \bar{X}^B + \frac{1}{2} (\im \cN)^{-1 \,AB} \big) \sum_x \big( \omega^x \otimes \omega^x \big)_{uv} \Big] \,. \label{vsimp}
\ee

\subsection{Tree level}\label{TreeLevel}
At tree level, the quaternionic vielbein is given by \cite{FerraraSabharwal}\footnote{$\varphi, \xi,\txi$ were introduced above. The coordinate $a$ is related to the dual $a_B$ of the spacetime component of the B-field via $a_B = a + \frac{\xi \txi}{2}$.} 
\ba 
u &=& \frac{1}{2} e^\varphi (d\txi - i d\xi)\,, \\
v &=& d\varphi - i\frac{e^{2 \varphi}}{2} \left(da + \txi d \xi \right)  \,.
\ea

The Sp(1) connection has the following form in terms of these quaternionic vielbein components\footnote{The components $\omega^x$ of the Sp(1) curvature $\omega$ should not be confused with the expansion forms $\omega_a$.} 
\be\label{eq:RelVielbeine-Connection}
\omega^1 = i (\bar{u} - u)\qquad,\qquad \omega^2 = -(u + \bar{u}) \qquad,\qquad \omega^3 = \frac{i}{2} (v - \bar{v}) \,.
\ee
In the class of theories we are considering, the isometries being gauged are described by the following Killing vectors
\bea\nnb
k_A &=& \sqrt{2} \left( e_{A} \frac{\partial}{\partial a} + q_{A} \frac{\partial}{\partial \xi} \right) \,, \\
\label{KillingVector}\tilde{k}^A &=& \sqrt{2} m^A \frac{\partial}{\partial a} \,.
\eea

Since $Q^u$ does not contain a non-vanishing entry for $u=\varphi$, the real part of $v$ does not enter upon contraction with $Q^u$, hence we can substitute
\ba
\sum_x \big( \omega^x \otimes \omega^x \big) &\sim& 4 u \otimes \bar{u} + v \otimes \bar{v}
\ea
in the potential, obtaining
\ba
V &=& Q_A^u \bar{Q}_B^v \Big[ - e^{2\varphi} \big(\frac{1}{2} (\im \cN)^{-1 \,AB}+3 e^K X^A \bar{X}^B  \big)  \big(d\xi^2 + d\txi^2 \big)_{uv} \\
&& \qquad \qquad \qquad \qquad \qquad \qquad \qquad \qquad-\frac{1}{8} e^{4 \varphi} (\im \cN)^{-1 \,AB} \big( da + \txi d \xi \big)^2_{uv}   \Big] \,.
\ea
This coincides with (\ref{eq:VNS}) and (\ref{eq:DefV_RR}) obtained above via reduction from 10 dimensions.

\subsection{All string loop} \label{allloop}

For the case of the universal hypermultiplet with 3 isometries, the quaternionic metric is of the Calderbank-Pedersen form \cite{Calderbank}. It comes in a 1-parameter family \cite{Strominger, AMTV}, determined by
\ban
u &=& \frac{\sqrt{\rho^2 + c}}{2(\rho^2- c)} (d\txi - id\xi) \,,\nn \\
v &=& -\frac{\rho}{2 (\rho^2 - c)\sqrt{\rho^2 + c}} \left[ 2 \frac{\rho^2 + c}{\rho} d \rho + i (d a + \txi d\xi)\right] \,. \label{uvcp}
\ean
The metric at string tree level lies at $c=0$, and the variable identification
\ba
\rho = e^{-\varphi} 
\ea
takes us back to the expression for the metric introduced above.\footnote{The coordinates used in \cite{AMTV} are related to our choice via $\psi = \frac{a+\xi \txi}{2} \,, \eta = -\frac{\xi}{2} \,,  \phi = \tilde{\xi}$.\label{cpc}}

In terms of the quaternionic vielbein components (\ref{uvcp}), the Sp(1) connection of the Calderbank-Pedersen metric is \cite{Calderbank}
\bea
\nnb\omega^1 &=& \frac{\rho}{\sqrt{\rho^2 + c}} i (\bar{u} - u) \,\,\,\,= \,\,\,-\frac{ \rho}{\rho^2-c} d\xi \,, \\ [2mm]
\nnb\omega^2 &=& -\frac{\rho}{\sqrt{\rho^2 + c}} (u + \bar{u})\,\, =\,\,\, - \frac{\rho}{\rho^2 - c} d \txi \,,\\ [2mm]
\label{eq:CPconnection}\omega^3 &=& \frac{\sqrt{\rho^2 + c}}{\rho} \frac{i}{2} (v- \bar{v}) \,\,\,=\,\, \frac{1}{2(\rho^2-c)} (da+\txi d\xi) \,.
\eea
The $\cN=2$ potential (\ref{vsimp}) for this choice of metric becomes
\ban
V &=& \frac{Q_A^u \bar{Q}_B^v}{(\rho^2-c)^2} \Big[ \big( -\frac{1}{2} (\im \cN)^{-1 \,AB} - 3 e^K X^A \bar{X}^B \big) \rho^2 (d\xi^2+d\txi^2)_{uv} \nn \\ [1mm]
&& -\frac{1}{8} (\im \cN)^{-1 \,AB} (da+\txi d\xi)^2_{uv} + c \,e^K  X^A \bar{X}^B (d\xi^2+d\txi^2)_{uv} \nn\\ [1mm]
&& - \frac{c}{\rho^2 + c}e^K X^A \bar{X}^B   (da+\txi d\xi)^2_{uv} \Big] \,. \label{potentialallloop}
\ean

In the case of Calabi-Yau compactifications, the metric is corrected away from $c = 0$ in passing from tree level to 1-loop \cite{AMTV}. Beyond 1-loop, all corrections can be captured by field redefinitions. This means that the quaternionic metric (i.e. the value of $c$) remains unchanged, the identification $\rho=e^{-\varphi}$ however is modified (note that the isometry structure of the metric determines the identification of the other 3 Calderbank-Pedersen coordinates with the 10d variables as indicated in footnote \ref{cpc}; this is why we have not introduced separate notation for them).

To study perturbative string corrections in the case of interest, let us review the argument of \cite{AMTV}. The 1-loop correction to the four-dimensional Einstein-Hilbert term can be determined by reduction of the 1-loop $R^4$ correction in 10d.\footnote{As with all such arguments, we are relying on the off-shell continuation of an on-shell string computation. It would be desirable to back this line of reasoning up with an explicit string computation on the background in question. We thank Pierre Vanhove for discussions on this point.} In the normalization of \cite{AMTV}, this yields
\ba
S_{{\rm Einstein-Hilbert}} &=& \int \, d^4 x \sqrt{g}\Big( e^{-2\varphi} -\frac{4 \zeta(2) \chi}{(2 \pi)^3}\Big)R \,.
\ea
Unfortunately, the full 1-loop corrected 10d action is not available as a means towards obtaining the 1-loop completion of the 4d action. Nonetheless, after parametrizing the ignorance regarding this action and comparing to the 4d effective action obtained by choosing the Calderbank-Pedersen metric on the universal hypermultiplet scalar manifold, \cite{AMTV} finds that only two possible values for $c$ are possible, 
\be\label{eq:Valuesc} c=0  \,\,\,\quad {\rm or}\,\,\,\quad c = - \frac{4 \zeta(2) \chi}{(2 \pi)^3}\,\,,\ee 
with $\chi$ the Euler characteristic of the Calabi-Yau. A perturbative string calculation then establishes that it is the latter value that is correct beyond tree level. Such a calculation in the case of the coset backgrounds with RR-flux that we are interested in is very challenging, and beyond the scope of this work. However, the first part of the analysis of \cite{AMTV} goes through also for these more general backgrounds. In particular, the 10d $R^4$ term is proportional to \cite{AMTV}
\ba
t_8 t_8 R^4 + \frac{1}{4} E_8 \,.
\ea
The first term is shorthand for $t_8 t_8 R^4 = t^{M_1 \cdots M_8} t^{N_1 \cdots N_8} R_{M_1 M_2 N_1 N_2} \cdots R_{M_7 M_8 N_7 N_8}$, which is expanded in terms of scalars built out of contractions of four Riemann tensors in eq. (A.12) of \cite{AMTV}. The second term can be written compactly in form notation as
\ba
E_8 \sim \Omega_{AB} \wedge \Omega_{CD} \wedge \Omega_{EF} \wedge \Omega_{GH} \wedge * (e^A \wedge\cdots \wedge e^H) \,,
\ea
with $\Omega^A{}_{B}=\frac{1}{2} R^A{}_{BCD}e^C e^D$ the curvature 2-form and $e^A$, $A=1,\ldots, 10$ a local coframe basis. From the expansion of the $t_8$ term in \cite{AMTV}, we see that in each scalar invariant, contractions pair at least two Riemann tensors. Hence, this term does not contribute to the 4d Einstein-Hilbert term upon reduction. The contribution from $E_8$ to the Einstein-Hilbert term stems, exactly as in the Ricci flat case, from
\ba
\Omega_{ab}\wedge *_4(e^a \wedge e^b) \wedge \Omega_{mn} \wedge \Omega_{pq} \wedge \Omega_{rs} \wedge *_6 (e^m \wedge \cdots \wedge e^s) \,,
\ea
with $a,b$ flat spacetime and $m, n, \ldots$ flat internal indices. We recognize the internal contribution as proportional to the 6 dimensional Euler density. The conclusion of our analysis is hence that in generalizing beyond Calabi-Yau manifolds, the same two possibilities for the Calderbank-Pedersen parameter $c$ exist as in the Calabi-Yau case (and await a perturbative string calculation as arbiter).

\section{Non-supersymmetric vacua} \label{nonsusy}
As an application of our consistent truncation result, we will search for non-supersymmetric vacua of the 4d effective action. By the analysis of section \ref{consistent}, these are guaranteed to lift to 10d solutions.

\subsection{Tree level} \label{nogo}
The potential we obtained at tree level above has the form
\ban
V&=& A_1 e^{2\varphi} + A_2 e^{4 \varphi} \,,  \label{pottree}
\ean
with
\ban 
A_1 &=& - Q_A^u \bar{Q}_B^v \big(\frac{1}{2} (\im \cN)^{-1 \,AB}+3 e^K X^A \bar{X}^B   \big)  \big( d\xi^2 + d\txi^2 \big)_{uv} \,, \nn \\
A_2 &=& - Q_A^u \bar{Q}_B^v \frac{1}{8} (\im \cN)^{-1 \,AB} \big( da + \txi d \xi \big)^2_{uv}  \,. \label{pottreeing}
\ean
Minimizing the potential with regard to the 4d dilaton yields \cite{AmirShamit}
\ba
V_\varphi = - \frac{A_1^2}{4A_2} \,.
\ea
As $A_2$ is positive definite, the potential at tree level is negative semi-definite on-shell. In fact, this result generalizes immediately to any hypermultiplet metric of the general form \cite{FerraraSabharwal} that arises upon Calabi-Yau and SU(3) structure compactifications, and the respective gaugings. The corresponding potential is obtained by appropriately modifying $u$ and $v$ in (\ref{pottreeing}). $A_2$ hence remains positive also in this more general case. 

We have thus proved that $\cN=2$ gauged supergravity as it arises in Calabi-Yau like compactifications at string tree level (i.e. with hypermultiplet metric as given in \cite{FerraraSabharwal}, and gaugings of axionic isometries) does not permit de Sitter solutions. Due to the consistency of the truncation, this 4d result also follows from the 10d no-go theorem of Maldacena-Nu\~nez \cite{MaldacenaNunez}. Note however that our 4d reasoning continues to hold for an {\it arbitrary} vector multiplet sector, i.e. including all possible worldsheet instanton corrections.

The two contributions to (\ref{pottree}) arise upon compactification from the NSNS and the RR sector respectively, see (\ref{eq:VNS}) and (\ref{eq:DefV_RR}). The positivity of $A_2$ is also manifest here.

\subsection{Non-supersymmetric Nearly K\"ahler companions}\label{companions}
The 10d analysis of subsection \ref{TheSolution} reveals that, given a choice of the RR fluxes $G_0$ and $G_6$, with all the other fluxes vanishing, there exists a single Nearly K\"ahler supersymmetric vacuum on the cosets (\ref{eq:OurCosets}). This solution is also recovered adopting the 4d approach, as discussed in \cite{HousePalti, KashaniPoorNearlyKahler}. 

It is possible to show that, under the same conditions, the 4d tree level scalar potential $V$ also admits non-supersymmetric Nearly K\"ahler extrema. In the following formulae, we introduce the sum of the geometric fluxes $q\equiv \sum_a q_a$, we rename the RR fluxes as $e_{0}\to e\,$, $m^0\to m$, and we call the equal $v^a$ and the equal $b^a$ respectively $v$ and $b$.

We obtain three Nearly K\"ahler extrema, lying at
\be \label{eq:SusycMin}
v = \frac{\sqrt{15}}{2}\left( \frac{1}{20I}\left|\frac{e}{m}\right|\right)^{1/3},\quad b = \frac{1}{2}\left(\frac{1}{20I} \frac{e}{m} \right)^{1/3},\quad  \tilde\xi= \frac{24 I m b^2}{q}  \;,\quad e^{2\varphi} = \frac{5 q^2}{48 I^2 m^{2} v^4} \;,
\ee
\be\label{eq:NonSusycMin}
v = \sqrt 3 \left( \frac{1}{20I}\left|\frac{e}{m}\right|\right)^{1/3},\quad b = -\left(\frac{1}{20I} \frac{e}{m} \right)^{1/3},\quad  \tilde\xi= -\frac{12 I m b^2}{q}  \;,\quad e^{2\varphi} =\frac{q^2 }{ 12 I^2 m^{2} v^4}\;,
\ee 
and
\be \label{eq:saddle}
v = \left( \frac{1}{\sqrt 5 I}\left|\frac{e}{ m}\right|\right)^{1/3}\;,\quad b\; = \;0 \;=\; \tilde\xi\;\;,\quad e^{2\varphi} = \frac{5q^2}{36 I^2 m^{2} v^4} \;.
\ee
By comparing to section \ref{TheSolution}, we learn that the only extremum preserving supersymmetry is (\ref{eq:SusycMin}).

Thanks to the consistency of the reduction, the non-supersymmetric extrema of $V$ found here also solve the 10d equations of motion, and actually turn out to coincide with the solutions previously found in ref. \cite{LustMarchMartTsimpis} via a 10d approach (see subsection 11.4 therein). 

Unlike the situation for the supersymmetric solution (\ref{eq:SusycMin}), for (\ref{eq:NonSusycMin}) and (\ref{eq:saddle}) stability is of course no longer guaranteed. As in any truncation scheme, a full stability analysis can only take place in the higher dimensional theory. What we can offer in our 4-dimensional theory is a stability analysis with regard to the modes we retain. To this end, we rescale the scalar fields\footnote{Note that the shift symmetry of $a$ and $\xi$ is gauged, the background value of these fields is hence a gauge choice.} $(v^a,b^a,\varphi,\txi)$ to obtain canonically normalized kinetic terms, and then diagonalize the mass matrix at the respective solutions.

\begin{figure}
\centering
\includegraphics[angle=0, scale=0.25]{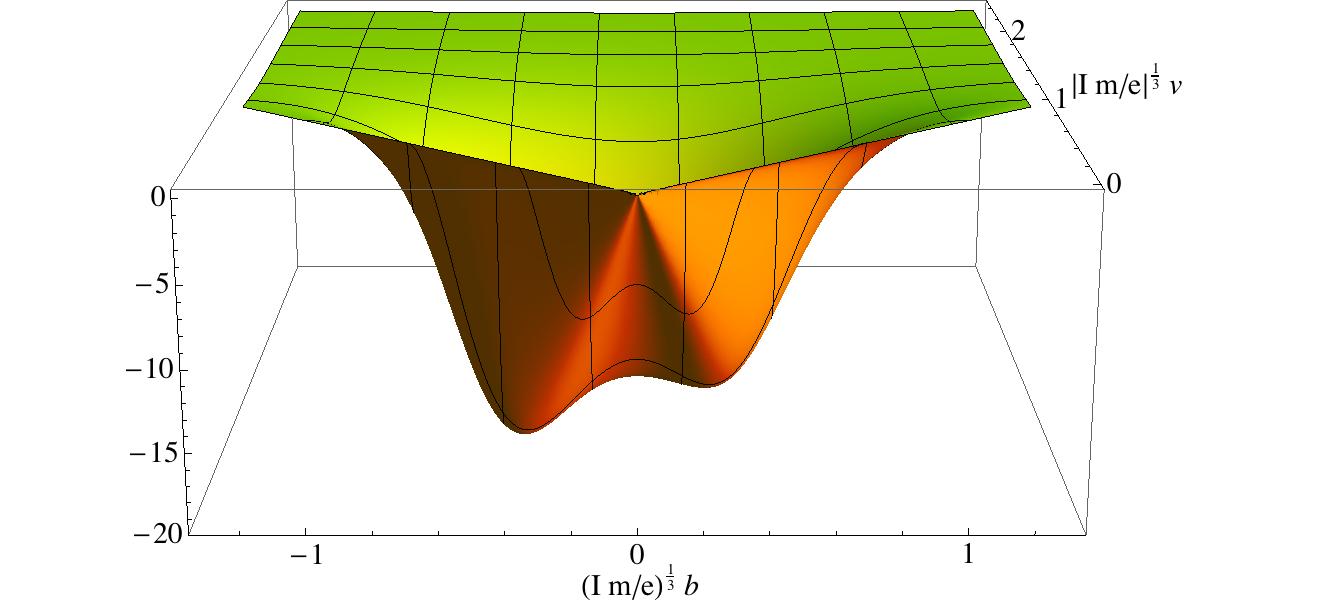}
\vskip -2mm
\caption{\footnotesize The potential for $\frac{\mathrm{G}_2}{\mathrm{SU(3)}\,}$: we plot the rescaled potential $e^{\frac{5}{3}}m^{\frac{1}{3}}I^{\frac{4}{3}}V$ as a function of $(Im/e)^{\frac{1}{3}}b\,$ and $\,|Im/e|^{\frac{1}{3}}v$, at the extremum of $\varphi$ and $\tilde\xi$. The deepest minimum corresponds to solution (\ref{eq:NonSusycMin}). The cut of the plot at $V=0$ is due to the constraint $e^{\varphi(b,v)}>0$.} \label{fig:potential}
\end{figure}

The case $\frac{\mathrm{G}_2}{\mathrm{SU(3)}}$ is depicted in figure \ref{fig:potential}: the first two extrema (\ref{eq:SusycMin}) and (\ref{eq:NonSusycMin}) are minima, while the remaining extremum is a saddle point. For $\frac{\mathrm{SU(3)}}{\mathrm{U(1)}\times \mathrm{U(1)}}$  and $\frac{\mathrm{Sp(2)}}{\mathrm{S(U(2)}\times\mathrm{U(1))}}$, (\ref{eq:NonSusycMin}) is a minimum, whereas due to modes leading away from the Nearly K\"ahler locus $v^a=v$ for all $a$, (\ref{eq:SusycMin}) is merely a saddle point, as is (\ref{eq:saddle}). To analyse stability, we compare the magnitude of the negative masses at the saddle points with the Breitenlohner-Freedman bound $$ m_{\rm tachyonic}^2 \ge - \frac{3}{4}|V| \,.$$ All extrema (including the saddle point depicted in figure \ref{fig:potential}) prove stable.

Finally, we remark that $\alpha'$ and string loop corrections can be safely neglected for the solutions above by tuning the RR fluxes $e$ and $m$ in such a way that the internal volume $V\!ol\equiv v^3I\sim e/m$ becomes sufficiently large and the string coupling constant $e^\phi\equiv e^\varphi \sqrt{V\!ol}\sim e^{-\frac{1}{6}}m^{-\frac{5}{6}}$ becomes small (recall the definition (\ref{eq:Def4dDilaton}) of the 4d dilaton). We can study moderately large string coupling by invoking the corrected potential (\ref{potentialallloop}). A numerical analysis indicates that all three AdS extrema survive string loop corrections. For the supersymmetric extremum, we push beyond numerics in appendix \ref{LoopCorrectionsToN=1}, and establish analytically that it persists, as expected, in the face of string loop corrections.

\subsection{de Sitter vacua at all string loop order?} \label{desitvac}
In face of the no-go result for de Sitter vacua obtained in subsection \ref{nogo}, we would like to analyse how loop corrections modify the outcome of this study. Of course, to guarantee the consistency of the truncation, the analysis in section \ref{consistent} must be extended beyond the two derivative case. However, the arguments put forth in subsection \ref{consistentint} in favor of consistency apply to the additional terms as well. We will also assume in this section that $c \neq 0$, as in the Calabi-Yau case. Note that by the results above, we can perform an (almost) complete analysis of the full loop corrected potential. The identification of the physical coordinate $\varphi$ and the Calderbank-Pedersen coordinate $\rho$, which is modified order by order in the string coupling and is not available, merely enters in identifying the range of the CP coordinate, see below. Away from very strong coupling (in which brane instanton corrections would have to be considered regardless), this does not affect the search for de Sitter minima. 

Focusing on the $\rho$ dependence of the potential (\ref{potentialallloop}) and taking the obvious positivity constraints on the coefficients into account does not rule out de Sitter vacua. One can then proceed to derive various constraints on these coefficients. E.g., by noting that the potential (\ref{potentialallloop}) has the form
\ba
V(\rho) &=& P(\rho) Q(\rho) \,,
\ea
with $P(\rho) = \frac{1}{(\rho^2 - c)^2}$, we obtain
\ba
V(\rho_0) &=& -\frac{P^2}{P'} Q' |_{\rho_0} \\
&=& \frac{Q_A^u \bar{Q}_B^v}{2 (\rho_0^2 - c)} \Big[ \big( -\frac{1}{2} (\im \cN)^{-1 \,AB} - 3 e^K X^A \bar{X}^B \big) (4  d\xi^2)_{uv}\\ && \hspace{4cm} + \frac{c}{(\rho_0^2 + c)^2}e^K X^A \bar{X}^B   (da + \txi d\xi)^2_{uv} \Big] \,,
\ea
where $\rho_0$ signifies the value of $\rho$ at a minimum of the potential.
Since $c$ is negative for the cosets we are considering, a de Sitter vacuum requires the first term in the square bracket to be positive at the minimum of the potential. This term is proportional to the tree level NSNS contribution to $V$, given in eq. (\ref{eq:VNS}). Hence, our necessary condition translates into the following inequality involving the internal NSNS 3--form and Ricci scalar
\be\nnb
H \lrcorner H - 2R_6 >0\,.
\ee
Recalling eq. (\ref{eq:RfromTorsion}), this is obviously true whenever the non-vanishing SU(3) torsion classes satisfy $15|W_1|^2 <  W_2\lrcorner \overline W_2$. For the simple case of Nearly K\"ahler manifolds (i.e. when $W_2=0$) the inequality is however non-trivial, and reads $3b^2-5v^2>0$.

We hope to return to a more complete analysis of the all loop corrected potential in the near future.

\section*{Acknowledgements}\vskip -2mm

We would like to thank Alessandro Tomasiello for collaboration in the initial stages of this project. We also acknowledge useful discussions with Adel Bilal, Paul Koerber, Luca Martucci, Ruben Minasian, Dimitrios Tsimpis and Pierre Vanhove. DC gratefully thanks the Service de Physique Th\'eorique et Math\'ematique de l'Universit\'e Libre de Bruxelles, where part of this work was done, for  hospitality and financial support. DC and AK thank the Erwin Schr\"odinger Institute in Vienna for hospitality during the ``Mathematical Challenges in String Phenomenology'' workshop.
DC and AK are supported in part by the EU grant MRTN-CT-2004-005104. In addition, DC is partially supported by the EU grant MRTN-CT-2004-512194, by the French grant ANR(CNRS-USAR) no.05-BLAN-0079-01 and by the ``Programme Vinci 2006 de l'Universit\'e Franco-Italienne''. AK is supported in part by {\it l'Agence  
Nationale de la Recherche} under the grants ANR-06-BLAN-3$\_$137168 and ANR-05-BLAN-0029-01.

\appendix

\section{Details of the dimensional reduction}\label{DetailsDimRed}

The $G$-invariant reduction ansatz strongly constrains the dependence of all the higher dimensional fields on the $G/H$ coordinates, relegating it into the coframe $e^{\ul m}$ introduced in subsection \ref{ExpBasis}. 
In particular, the most general $G$-invariant 10d metric is (here and in the following, the hat denotes 10d fields):
\be\label{eq:10dMetricAnsatzCoset}
d \hat s^2 =  e^{2\varphi(x)}g_{\mu\nu}(x)dx^\mu\otimes dx^\nu + g_{\ul m\ul n}(x)e^{\ul m}(y)\otimes e^{\ul n}(y)\;,
\ee
where $x^\mu$ and $y^m$ are respectively coordinates on the 4d spacetime and the internal manifold $M_6$, and $g_{\ul m\ul n}$ satisfies the $G$-invariance condition discussed in subsection \ref{ExpBasis}. Components of the 10d metric with mixed 4d-6d indices are not allowed since there are no left-invariant 1--forms on our coset manifolds (\ref{eq:OurCosets}). Since the invariant scalars on the coset are necessarily constant, a non-trivial warp factor is also not permitted (see \cite{KoerberMartucci,ShiuTUDouglas} for recent discussions of a non-trivial warp factor in the $\cN=1$ context). The Weyl rescaling factor $e^{2\varphi(x)}$ in front of the 4d metric is needed in order to obtain a canonical lower dimensional Einstein-Hilbert term $\int_{M_4}vol_4 R_4$ from the string frame higher dimensional action $\int_{M_{10}}vol_{10} e^{-2\phi} \hat R$, with 
\be\label{eq:Def4dDilaton}
\varphi(x)= \phi(x) - \frac{1}{2}\log\int_{M_6} d^6y \sqrt{g_{6}}\;,
\ee
where $\phi(x)$ is the 10d dilaton and $\sqrt{g_6} \equiv\sqrt{\det g_{mn}(x,y)}=  \sqrt{\det g_{\ul m\ul n}(x)}\,|\det e^{\ul p}_{\;\,q}(y)|\,$. Notice that, thanks to this factorization of the $x$ and $y$ dependence, $\partial_\mu\log\sqrt{g_6}$ does not depend on the internal coordinates, and
\be\label{eq:DerivativeVarphi}
\partial_\mu\varphi = \partial_\mu\phi - \frac{1}{2}\partial_\mu\log \sqrt{g_6} \;.
\ee
The ansatz for the 10d supergravity field strengths must be chosen consistently with their Bianchi identities. For instance, from the Bianchi identity $d \hat F_2= \hat H \hat F_0$, one sees that if $\hat F_0 \neq 0$, then the NSNS 3--form $\hat H$ has to be exact: $\hat H= d\hat B$, with a globally defined 2--form potential $\hat B$. The most general $\hat B$ respecting left-invariance on $M_6$ is
\be\label{eq:SplitB}
\hat B= B + b\;,
\ee 
where $B(x)$ is along 4d spacetime, while $b(x,y) = b^a(x)\om_a(y)$ lives on $M_6$ (the left-invariant 2--forms $\om_a$ were given in subsection$\:$\ref{ExpBasis}). 

We deal with the expansion of the RR fields in subsection$\:$\ref{ReductionRRsector}.

\subsection{Special K\"ahler geometry from the NSNS sector}\label{SpKahlerGeo}

Combining the 2--form $J$ of subsection \ref{SU3str} and the internal NS field $b$ we introduce $t = b+ iJ $, whose expansion $t=t^a\om_a$ on the basis 2--forms defines the complex 4d scalars $t^a=b^a + iv^a$. The associated kinetic term is determined by
\be\label{eq:KineticTermsNSNSfields}
\frac{1}{8} g^{mp}g^{nq}\big( \partial_\mu g_{m n}\partial^\mu g_{pq} +  \partial_\mu b_{mn }\partial^\mu b_{pq} \big) \;=\; \frac{1}{4V\!ol}\int_{M_6} \partial_\mu t\wedge * \partial^\mu \bar t  \;=\; \cl G_{ab}\partial_\mu t^a \partial^\mu \bar t^b\;,
\ee
where the l.h.s. originates from the reduction of the 10d Ricci scalar and $\hat H^2$ terms, while the $\sigma$-model metric $\cl G_{ab}$ was introduced in eq.$\:$(\ref{eq:Gfrom2forms}). The first equality in (\ref{eq:KineticTermsNSNSfields}) is derived recalling that the internal metric is fixed by the forms $J$ and $\Om$ defining the SU(3) structure: indeed, calling $\cl I$ the almost complex structure induced by $\Om$, we have $g_{mn}= J_{mp}\cl I^p_{\;\;n}$. Notice that we get no contribution from the variation of $\cl I$ since the associated $\Om$, given in eq.$\:$(\ref{eq:JandOmForCoset}), is rigid.

The metric $\cl G_{ab}$ is special K\"ahler: indeed, it can be obtained via $\cl G_{ab} = \frac{\partial^2K}{\partial t^a\partial \bar t^b}$ from the K\"ahler potential
\be\label{eq:KahlerPotential}
K = -\log \frac{4}{3}\int J\wedge J\wedge J \;=\;-\log 8V\!ol\;.
\ee
It in turn is determined by a prepotential $\cl F$ via the special K\"ahler geometry formula $K = -\log i(\,\overline X^A\cl F_A - X^A\overline{\cl F}_A\,)$, where $X^A\equiv (X^0,X^a)=(1,-t^a)$ and ${\cl F}_A=\frac{\partial \cl F(X)}{\partial X^A}$. 

For each of the cosets we consider, the explicit expressions of $\cl G_{ab}$ and $V\!ol$ were given in table \ref{DetailsCosets}. The (cubic) prepotential reads\vskip -2mm
\be\nnb
\cl F(X) = \frac{1}{6}\cl K_{abc} \frac{X^aX^bX^c}{X^0}\;,
\ee
where the non-vanishing triple intersection numbers $\cl K_{abc}:=\int \om_a\wedge\om_b\wedge\om_c$ (recall the 2--forms $\om_a$ in subsection \ref{ExpBasis}) are\vskip -2mm
\be\label{eq:IntNumb} \begin{array}{ll}
\cl K_{123}\, =\, \;I &\quad\textrm{for}\quad\frac{\mathrm{SU(3)}}{\mathrm{U(1)}\times \mathrm{U(1)}}\\ [2mm]
\cl K_{112}\, =\, 2I &\quad\textrm{for}\quad\frac{\mathrm{Sp(2)}}{\mathrm{S(U(2)}\times\mathrm{U(1))}}\\ [2mm]
\cl K_{111}\, =\, 6I &\quad\textrm{for}\quad\frac{\mathrm{G}_2}{\mathrm{SU(3)}}\;.
\end{array}
\ee
The period matrix $\cl N_{AB}$ of special K\"ahler geometry is given by the formula (see e.g.$\:$\cite{WhatIsSpecialKaehler?})
\be\nnb
\cl N_{AB} = \overline{\cl F}_{AB} + 2i \frac{\im (\cl F_{AC})X^C\im (\cl F_{BD})X^D}{X^E\im (\cl F_{EF})X^F}\;,\quad\textrm{where }\; \cl F_{AB}\equiv\frac{\partial^2\cl F}{\partial X^A\partial X^B}\;.
\ee
Equivalently, we can directly obtain it from the coset geometry via \cite{CassaniBilal}:
\bea
\nnb (\im\cl N)^{-1\,AB} = -\int \langle\tilde\om^A , *_b\tilde\om^B\rangle\quad &,&\quad  [\re\cl N(\im\cl N)^{-1}]_A^{\;B} = - \int \langle\om_A, *_b\tilde\om^B\rangle\;,\\
\nnb [\im\cl N +\re\cl N(\im\cl N)^{-1}\re\cl N]_{AB} &=& -\int\langle \om_A, *_b\om_B\rangle\;,
\eea
with $\,*_b(\,\cdot\,)\,\equiv\,e^{-b}\,*\,\lambda (e^b\,\cdot\, )\,\,$. The operator $\lambda$ and the pairing $\langle\,,\,\rangle$ were defined below$\;$(\ref{eq:PairingBasisForms}). 

We obtain the matrices
\be\label{eq:ImN}
\im\mathcal{N} \;=\; -V\!ol \left( \begin{array}{cc}
1+4\cl G_{ab}b^ab^b \;&\; 4\cl G_{ab} b^b\\ [2mm]
4\cl G_{ab} b^b \;&\;  4\cl G_{ab}
\end{array} \right)\;,
\ee
\be\label{eq:ReN}
\re\cl N\; =\; -\left(\begin{array}{cccc}
\frac{1}{3}\cl K_{abc}b^ab^bb^c \;&\; \frac{1}{2}\cl K_{abc}b^bb^c \\ [2mm]
\frac{1}{2}\cl K_{abc}b^bb^c \;&\;   \cl K_{abc} b^c 
\end{array} \right)\;.
\ee

\subsection{The RR sector}\label{ReductionRRsector}

In order to reduce the RR sector we specialize the general procedure described in section$\:$5 of ref.$\:$\cite{ReducingSU3SU3} for $M_6$ corresponding to our coset spaces. Adopting the democratic formulation of type IIA supergravity \cite{Democratics}, the RR degrees of freedom can be encoded in a field strength ${\bf \hat G} $ consisting of a formal sum of forms of all possible even degrees, satisfying
\bea
\label{eq:Bianchi10dG} \!\!\!\!\!\!\!\!\!\!\!\!\!\!\!\textrm{Bianchi identity }\!\!\!\!&:& \; d{\bf \hat G} = 0  \\ [1mm]
\label{eq:SelfDuality10dG}\!\!\!\!\!\!\!\!\!\!\!\!\!\!\!\textrm{self-duality constraint}\!\!\!\!&:&\; {\bf \hat F} =  \lambda (* {\bf \hat F})\,,\;\;\; \textrm{where } {\bf \hat F} \equiv e^{\hat B}{\bf \hat G}\;\textrm{and}\;\lambda(\hat F_{(k)}) = (-)^{\frac{k}{2}}\hat F_{(k)}.
\eea
Due to the self-duality constraint, the equations of motion for the RR degrees of freedom are equivalent to the Bianchi identities.

We implement the reduction ansatz by expanding ${\bf \hat{G}}$ on the basis of left-invariant internal forms introduced in subsection \ref{ExpBasis},
\be\label{eq:ExpansionWithG}
{\bf \hat{G}} = (G_{(0)}^A + G_{(2)}^A + G_{(4)}^A) \om_A - (\tilde G_{(0)A} + \tilde G_{(2)A} + \tilde  G_{(4)A}) \tilde \om^A + (G_{(1)} + G_{(3)})\alpha - (\tilde G_{(1)} + \tilde G_{(3)})\beta.
\ee
$G_{(p)}(x)$ and $\tilde G_{(p)}(x)$ are $p$--forms in 4d spacetime. Plugging this expansion into eqs.$\:$(\ref{eq:Bianchi10dG}), (\ref{eq:SelfDuality10dG}), and going through the derivation of \cite{ReducingSU3SU3}, one identifies the 4d variables\vskip -7mm
\bea\label{eq:4dDOF}
 G_{(0)}^A = m^A \quad &,& \quad \tilde G_{(0)A}= e_{A} + q_A \,\tilde\xi \\ [2mm]
\nnb G_{(1)} = D\xi \equiv d\xi -  q_a A^a\quad &,&\quad \tilde G_{(1)}=d\tilde\xi \\ [2mm]
\nnb G_{(2)}^A = dA^A\quad &,&\quad \tilde G_{(2)A} + B \tilde G_{(0)A} = \im \mathcal N_{AB} *F^B + \re\mathcal N_{AB} F^B\\ [2mm]
\nnb G_{(3)} =- B \wedge D\xi + e^{2\varphi}* d\tilde \xi \quad &,& \quad \tilde G_{(3)} = - B \wedge d\tilde \xi -e^{2\varphi}* D\xi
\eea 
\be
\nnb G_{(4)}^A + B\wedge G_{(2)}^A+ \frac{1}{2} B^2\, G_{(0)}^A = e^{4\varphi}\big[(\im\cl N)^{-1}(\tilde G_{(0)}-\re{\cl N}G_{(0)})\big]^A *1
\ee
\be
\nnb \tilde G_{(4)A} + B\!\wedge\! \tilde G_{(2)A}+\frac{1}{2} B^2\tilde G_{(0)A} = e^{4\varphi}\big[\!-\im\!\cl N G_{(0)}+ \re \cl N(\im\cl N)^{-1}(\tilde G_{(0)}-\re{\cl N}G_{(0)}) \big]_A\!\!*1
\ee
where the propagating fields are the two real scalars $\xi,\tilde \xi$ and the 1--forms $A^A$. We also introduced the modified field strengths
\be\label{eq:ModifFieldStrp=0=qCase}
F^A \,\equiv\, dA^A +  m^A B\;.
\ee
Furthermore we introduce $q_A=(0,q_a)$, the $q_a$ being the geometric fluxes defined in subsection \ref{properties}, while $m^A , e_{A}$ are constant flux parameters satisfying $q_a m^a = 0$. Notice that one of the $e_a$ is redundant, since it can be eliminated via a constant shift of $\tilde\xi\,$. This reflects the fact that on our cosets the linear combination $q_a\tilde\om^a$ is exact (see eq. (\ref{diffrel})), and therefore doesn't support any flux. 

The residual content of (\ref{eq:Bianchi10dG})--(\ref{eq:ExpansionWithG}) not included in eqs.$\:$(\ref{eq:4dDOF}) consists of a set of equations to be read as the EoM for $\xi,\tilde\xi$ and $A^A$. We use these equations to reconstruct the 4d action $S^{(4)}_{\mathrm{RR}}$ of subsection$\:$\ref{4daction}. In particular, we infer the RR contribution to the 4d scalar potential,
\be\label{eq:V_RRgeneral}
V_{\mathrm{RR}} = -\frac{e^{4\varphi}}{ 4}\big[ G_{(0)}\im\mathcal{N}G_{(0)} + (\tilde G_{(0)} - G_{(0)}\re\mathcal{N})(\im\mathcal{N})^{-1}(\tilde G_{(0)} - \re\mathcal{N}G_{(0)}) \big]\;.
\ee
Substitution of the explicit expressions for $G_{(0)}$ and $\tilde G_{(0)}$ given in (\ref{eq:4dDOF}) yields eq. (\ref{eq:DefV_RR}).

As a last remark, we stress that the whole procedure of section~5 of \cite{ReducingSU3SU3} applies here with no need to take any integral over $M_6$. In other words, once the left-invariant truncation ansatz has been plugged in, the dependence of eqs.$\:$(\ref{eq:Bianchi10dG}), (\ref{eq:SelfDuality10dG}) on the internal coordinates automatically factorizes out.

\section{String loop corrections to the $\cl N=1$ vacua}\label{LoopCorrectionsToN=1}

In this appendix, we study how string loop corrections affect the tree level supersymmetric AdS$_4$ solutions of type IIA supergravity compactified on the cosets (\ref{eq:OurCosets}).  

The $\cl N=1$ equations arise by requiring the vanishing of the fermionic (i.e. gravitino-, hyperino- and gaugino-) variations under a single linear combination of the two $\cl N=2$ susy parameters. These conditions have been spelled out in \cite{CassaniBilal} for general SU(3)$\times$SU(3) structure compactifications, and solved in \cite{KashaniPoorNearlyKahler} for the subclass of Nearly K\"ahler manifolds. Here, we extend the latter analysis employing the string loop corrected quaternionic vielbein (\ref{uvcp}) and the associated Sp(1) connection. In particular, the Killing prepotentials associated with our electric and magnetic gaugings of the quaternionic isometries become, recalling relation (\ref{eq:RelPomega}), the Killing vectors (\ref{KillingVector}), and the Calderbank-Pedersen Sp(1) connection (\ref{eq:CPconnection}), 
\bea\nnb
\cl P^1_A= -\frac{\sqrt{2}\rho}{\rho^2-c}q_A\quad &,&\quad \tilde{\cl P}^{1A} = \cl P^2_A = \tilde{\cl P}^{2A} = 0\,,\\
\label{eq:KillPrepExplicit}\cl P^3_A = \frac{\sqrt{2}}{2(\rho^2-c)}(e_A + \tilde\xi q_A)\quad&,&\quad \tilde {\cl P}^{3A}= \frac{\sqrt{2}}{2(\rho^2-c)} m^A\,.
\eea
The tree level Killing prepotentials are recovered by taking $c=0$ (recall the possible values of $c$, given in (\ref{eq:Valuesc})), together with the identification $\rho^2 = e^{-2\varphi}$. The first part of the analysis performed in subsection 6.1 of \cite{KashaniPoorNearlyKahler} goes through in the present case, the only substantial difference being that the relation between the quaternionic vielbein $u,v$ and the Sp(1) connection $\om^x$ is here slightly more involved than (\ref{eq:RelVielbeine-Connection}); this leads to a modification of the equations arising from the hyperino variation. After a few manipulations, we arrive at the following $\cl N=1$ AdS vacuum condition for our coset reductions (both $\pm$ signs are allowed by susy),
%
%
\be\label{eq:GauginoVar}
-\left[(\im \cl N)^{-1\,AB} +  \frac{3\rho^2 +c}{\rho^2}e^K  \bar X^A X^B\right]\cl P^1_B \,\pm\, i(\im \cl N)^{-1\,AB}\big (\cl P^3_B - \cl N_{BC }\cl {\tilde P}^{3C}\big) \,=\,0\;,
\ee
the (string frame) AdS cosmological constant being given by
\be\label{eq:AdScurvSusyVacuum}
\Lambda = -\frac{3}{2}e^K |q_A X^A|^2\;.
\ee
We now solve the susy condition in the Nearly K\"ahler limit. As in subsection \ref{companions},  we define $q\equiv \sum_a q_a$, we rename the only non-vanishing fluxes as $e_{0}\to e\,$, $m^0\to m$, and we set $v^a=v$ and $b^a=b$ for all $a$. Separating (\ref{eq:GauginoVar}) into real and imaginary parts, and recalling (\ref{eq:KahlerPotential}) for $K$, as well as (\ref{eq:ImN}), (\ref{eq:ReN}) for $\cl N$, we obtain the four real equations
\bea\nnb
b = \pm \frac{4\rho}{5\rho^2-c}\frac{mIv^3}{q}\quad &,&\quad b^2 = \frac{\rho^2+ 3c}{15\rho^2-3c}v^2\; \\ [2mm]
\nnb  -be + \big(b^2 + \frac{v^2}{3}\big)q\tilde\xi+ mI(b^4 +v^2b^2)=0\quad &,& \quad -e + bq\tilde\xi +  m I b^3 \pm \frac{3\rho^2+c}{4\rho}qv=0\;.
\eea
This system of equations is solved by
\be\label{eq:SolLoopCorrSusy}
v = v_T \frac{5x-\tilde c}{(5x+3\tilde c)x^{\frac{1}{2}}}\quad,\quad b=b_T \left[\frac{(x+3\tilde c)^3}{x(x-\tilde c/5)}\right]^{\frac{1}{4}}\;\;,\quad\tilde\xi = \xi_T \left[\frac{x(x+ 3 \tilde c)}{x - \tilde c/5} \right]^{\frac{1}{2}}\;\;,\quad  \rho^2 = \rho_T^2x\;,\ee
where by $v_T, b_T, \tilde \xi_T, \rho_T^2$ we denote the tree level values (\ref{eq:SusycMin}) (recall that at tree level $\rho^2$ is identified with $e^{-2\varphi}$). We have also defined $\tilde c\! = \!\rho_T^{-2} c$ (note that this depends on the values of the fluxes appearing in $\rho_T^2\sim (me^2I)^{\frac{2}{3}}q^{-2}$), while $x$ is the unique positive solution to the equation
\be\label{eq:EqForx}
(5x + 3\tilde c)^4(x + 3\tilde c)x - 5 (5x - \tilde c)^3 = 0\,,
\ee
and can easily be determined numerically. The cosmological constant (\ref{eq:AdScurvSusyVacuum}) here reads 
\be\nnb
\Lambda = -\frac{q^2}{5Iv_T}\frac{ (x+3\tilde c/5)x^{\frac{3}{2}} }{(x-\tilde c/5)^2}\;.
\ee

The tree level result is recovered by taking $\tilde c=0$, in which case (\ref{eq:EqForx}) is solved by $x=1$. We have also checked that (\ref{eq:SolLoopCorrSusy}), (\ref{eq:EqForx}) extremize the all loop scalar potential (\ref{potentialallloop}).

We conclude that string loops preserve the main outcome of the tree level analysis: for any choice of the fluxes $e,m$, there exists a unique Nearly K\"ahler supersymmetric solution. This is however shifted from the tree level position as shown in (\ref{eq:SolLoopCorrSusy}). It would be interesting to study the lifting of this result to a 10d framework.



\begin{thebibliography}{99}

\bibitem{Louis}
  M.~Gomez-Reino, J.~Louis and C.~A.~Scrucca,
  {\it No metastable de Sitter vacua in N=2 supergravity with only
  hypermultiplets},
  arXiv:0812.0884 [hep-th].

\bibitem{Vandoren}
  F.~Saueressig, U.~Theis and S.~Vandoren,
  {\it On de Sitter vacua in type IIA orientifold compactifications},
  Phys.\ Lett.\  B {\bf 633}, 125 (2006)
  [arXiv:hep-th/0506181].

\bibitem{BodnerCadavidFerrara}
  M.~Bodner, A.~C.~Cadavid and S.~Ferrara,
  {\it (2,2) vacuum configurations for type IIA superstrings: N=2 supergravity
  Lagrangians and algebraic geometry},
  Class.\ Quant.\ Grav.\  {\bf 8} (1991) 789.

\bibitem{BohmGHLouis}
  R.~Bohm, H.~Gunther, C.~Herrmann and J.~Louis,
  {\it Compactification of type IIB string theory on Calabi-Yau threefolds},
  Nucl.\ Phys.\  B {\bf 569} (2000) 229
  [arXiv:hep-th/9908007].

\bibitem{generalized mirror symmetry}
  S.~Gurrieri, J.~Louis, A.~Micu and D.~Waldram,
  {\it Mirror symmetry in generalized Calabi-Yau compactifications},
  Nucl.\ Phys.\  B {\bf 654}, 61 (2003)
  [arXiv:hep-th/0211102].
  
\bibitem{GiddingsKachruPolchinski}
  S.~B.~Giddings, S.~Kachru and J.~Polchinski,
  {\it Hierarchies from fluxes in string compactifications},
  Phys.\ Rev.\  D {\bf 66} (2002) 106006
  [arXiv:hep-th/0105097].

\bibitem{GMPT1}
  M.~Grana, R.~Minasian, M.~Petrini and A.~Tomasiello,
  {\it Supersymmetric backgrounds from generalized Calabi-Yau manifolds},
  JHEP {\bf 0408} (2004) 046
  [arXiv:hep-th/0406137].

\bibitem{GMPT2}
  M.~Grana, R.~Minasian, M.~Petrini and A.~Tomasiello,
  {\it Generalized structures of N=1 vacua},
  JHEP {\bf 0511} (2005) 020
  [arXiv:hep-th/0505212].

\bibitem{PolchinskiStrominger}
  J.~Polchinski and A.~Strominger,
  {\it New Vacua for Type II String Theory},
  Phys.\ Lett.\  B {\bf 388} (1996) 736
  [arXiv:hep-th/9510227].

\bibitem{N=2review}
L.~Andrianopoli, M.~Bertolini, A.~Ceresole, R.~D'Auria, S.~Ferrara, P.~Fre and T.~Magri,
  {\it $N = 2$ supergravity and $N = 2$ super Yang-Mills theory on general scalar
  manifolds: Symplectic covariance, gaugings and the momentum map},
  J.\ Geom.\ Phys.\  {\bf 23}, 111 (1997)
  [arXiv:hep-th/9605032].

\bibitem{ReducingSU3SU3} 
  D.~Cassani,
  {\it Reducing democratic type II supergravity on SU(3) $\times$ SU(3) structures},
  JHEP {\bf 0806} (2008) 027
  [arXiv:0804.0595 [hep-th]].

\bibitem{KashaniPoorNearlyKahler}
  A.~K.~Kashani-Poor,
  {\it Nearly Kaehler Reduction},
  JHEP {\bf 0711} (2007) 026
  [arXiv:0709.4482 [hep-th]].

\bibitem{DuffNilssonPopeKKreview}
  M.~J.~Duff, B.~E.~W.~Nilsson and C.~N.~Pope,
  {\it Kaluza-Klein Supergravity},
  Phys.\ Rept.\  {\bf 130} (1986) 1.
  
\bibitem{KashaniPoorMinasian}
	A.~K.~Kashani-Poor and R.~Minasian,
  {\it Towards reduction of type II theories on SU(3) structure manifolds},
  JHEP {\bf 0703}, 109 (2007)
  [arXiv:hep-th/0611106].

\bibitem{GrossWitten}
  D.~J.~Gross and E.~Witten,
  {\it Superstring Modifications Of Einstein's Equations},
  Nucl.\ Phys.\  B {\bf 277} (1986) 1.

\bibitem{GreenSchwarz}
  M.~B.~Green and J.~H.~Schwarz,
  {\it Supersymmetrical Dual String Theory. 3. Loops And Renormalization},
  Nucl.\ Phys.\  B {\bf 198} (1982) 441.
  
\bibitem{SakaiTanii}
  N.~Sakai and Y.~Tanii,
  {\it One Loop Amplitudes And Effective Action In Superstring Theories},
  Nucl.\ Phys.\  B {\bf 287} (1987) 457.

\bibitem{PolicastroTsimpis}
  G.~Policastro and D.~Tsimpis,
  {\it R**4, purified},
  Class.\ Quant.\ Grav.\  {\bf 23} (2006) 4753
  [arXiv:hep-th/0603165].

\bibitem{PeetersVW}
  K.~Peeters, P.~Vanhove and A.~Westerberg,
  {\it Supersymmetric higher-derivative actions in ten and eleven dimensions, the
  associated superalgebras and their formulation in superspace},
  Class.\ Quant.\ Grav.\  {\bf 18} (2001) 843
  [arXiv:hep-th/0010167].

\bibitem{CassaniBilal}
D.~Cassani and A.~Bilal,
  {\it Effective actions and N=1 vacuum conditions from SU(3)$\times$SU(3) compactifications},
  JHEP {\bf 0709} (2007) 076
  [arXiv:0707.3125 [hep-th]].

\bibitem{Hertzberg}
  M.~P.~Hertzberg, S.~Kachru, W.~Taylor and M.~Tegmark,
  {\it Inflationary Constraints on Type IIA String Theory},
  JHEP {\bf 0712} (2007) 095
  [arXiv:0711.2512 [hep-th]].

\bibitem{MinimalSimpledS}
  S.~S.~Haque, G.~Shiu, B.~Underwood and T.~Van Riet,
  {\it Minimal simple de Sitter solutions},
  arXiv:0810.5328 [hep-th].
  
\bibitem{CosmologySU3}
  C.~Caviezel, P.~Koerber, S.~Kors, D.~Lust, T.~Wrase and M.~Zagermann,
  {\it On the Cosmology of Type IIA Compactifications on SU(3)-structure
  Manifolds},
  arXiv:0812.3551 [hep-th].
    
\bibitem{FlaugerPabanRobbinsWrase}
  R.~Flauger, S.~Paban, D.~Robbins and T.~Wrase,
  {\it On Slow-roll Moduli Inflation in Massive IIA Supergravity with Metric Fluxes},
  arXiv:0812.3886 [hep-th].
  
\bibitem{Neupane}
  I.~P.~Neupane,
   {\it Accelerating universes from compactification on a warped conifold},
  Phys.\ Rev.\ Lett.\  {\bf 98} (2007) 061301
  [arXiv:hep-th/0609086];
  {\it Simple cosmological de Sitter solutions on dS$_4 \times Y_6$ spaces},
  arXiv:0901.2568 [hep-th].

\bibitem{KoerberLustTsimpis}
  P.~Koerber, D.~Lust and D.~Tsimpis,
  {\it Type IIA AdS4 compactifications on cosets, interpolations and domain walls},
  JHEP {\bf 0807}, 017 (2008)
  [arXiv:0804.0614 [hep-th]].

\bibitem{GranaReview}
  M.~Grana,
  {\it Flux compactifications in string theory: A comprehensive review},
  Phys.\ Rept.\  {\bf 423} (2006) 91
  [arXiv:hep-th/0509003].

\bibitem{ChiossiSalamon}
  S.~Chiossi and S.~Salamon,
  {\it The intrinsic torsion of SU(3) and G$_2$ structures},
  arXiv:math/0202282.

\bibitem{MuellerStuckl}
  F.~Mueller-Hoissen and R.~Stuckl,
  {\it Coset spaces and ten-dimensional unified theories},
  Class.\ Quant.\ Grav.\  {\bf 5} (1988) 27.
  
\bibitem{BehrndtCveticShort}
  K.~Behrndt and M.~Cvetic,
  {\it General N = 1 Supersymmetric Flux Vacua of (Massive) Type IIA String
  Theory},
  Phys.\ Rev.\ Lett.\  {\bf 95} (2005) 021601
  [arXiv:hep-th/0403049].

\bibitem{HousePalti}
  T.~House and E.~Palti,
  {\it Effective action of (massive) IIA on manifolds with SU(3) structure},
  Phys.\ Rev.\  D {\bf 72}, 026004 (2005)
  [arXiv:hep-th/0505177].

\bibitem{TomasielloTwistor}
  A.~Tomasiello,
  {\it New string vacua from twistor spaces},
  Phys.\ Rev.\  D {\bf 78} (2008) 046007
  [arXiv:0712.1396 [hep-th]].

\bibitem{AldazabalFont}
  G.~Aldazabal and A.~Font,
  {\it A second look at $\cl N=1$ supersymmetric AdS$_4$ vacua of type IIA supergravity},
  JHEP {\bf 0802} (2008) 086
  [arXiv:0712.1021 [hep-th]].

\bibitem{CaviezelKoerberKorsLustTsimpisZagermann}
  C.~Caviezel, P.~Koerber, S.~Kors, D.~Lust, D.~Tsimpis and M.~Zagermann,
  {\it The effective theory of type IIA AdS4 compactifications on nilmanifolds and
  cosets},
  Class.\ Quant.\ Grav.\  {\bf 26}, 025014 (2009)
  [arXiv:0806.3458 [hep-th]].

\bibitem{Lust1}
  D.~Lust,
  {\it ``Compactification Of Ten-Dimensional Superstring Theories Over Ricci Flat Coset Spaces},
  Nucl.\ Phys.\  B {\bf 276} (1986) 220.

\bibitem{CastellaniLust}
  L.~Castellani and D.~Lust,
  {\it Superstring compactification on homogeneous coset spaces with torsion},
  Nucl.\ Phys.\  B {\bf 296} (1988) 143.

\bibitem{HeteroticOnCosets}
  A.~Chatzistavrakidis, P.~Manousselis and G.~Zoupanos,
  {\it Reducing the Heterotic Supergravity on nearly-Kahler coset spaces}
  arXiv:0811.2182 [hep-th].

\bibitem{MooreWitten}
  G.~W.~Moore and E.~Witten,
  {\it Self-duality, Ramond-Ramond fields, and K-theory},
  JHEP {\bf 0005} (2000) 032
  [arXiv:hep-th/9912279].

\bibitem{MinasianMoore}
  R.~Minasian and G.~W.~Moore,
  {\it ``K-theory and Ramond-Ramond charge},
  JHEP {\bf 9711} (1997) 002
  [arXiv:hep-th/9710230].

\bibitem{DMW}
  D.~E.~Diaconescu, G.~W.~Moore and E.~Witten,
  {\it E(8) gauge theory, and a derivation of K-theory from M-theory},
  Adv.\ Theor.\ Math.\ Phys.\  {\bf 6} (2003) 1031
  [arXiv:hep-th/0005090].

\bibitem{Karoubi}
 M.~Karoubi, {\it $K$-theory}, Springer Verlag, 2008.

\bibitem{LustTsimpis}
  D.~Lust and D.~Tsimpis,
  {\it Supersymmetric AdS(4) compactifications of IIA supergravity},
  JHEP {\bf 0502} (2005) 027
  [arXiv:hep-th/0412250].

\bibitem{GMPT3}
  M.~Grana, R.~Minasian, M.~Petrini and A.~Tomasiello,
  {\it ``A scan for new N=1 vacua on twisted tori},
  JHEP {\bf 0705} (2007) 031
  [arXiv:hep-th/0609124].

\bibitem{Democratics}
  E.~Bergshoeff, R.~Kallosh, T.~Ortin, D.~Roest and A.~Van Proeyen,
  {\it New formulations of $D = 10$ supersymmetry and $D8 - O8$ domain walls},
  Class.\ Quant.\ Grav.\  {\bf 18} (2001) 3359
  [arXiv:hep-th/0103233].

\bibitem{CastellaniRomansWarner}
  L.~Castellani, L.~J.~Romans and N.~P.~Warner,
  {\it Symmetries Of Coset Spaces And Kaluza-Klein Supergravity},
  Annals Phys.\  {\bf 157} (1984) 394.

\bibitem{DuffNilssonPopeWarner}
  M.~J.~Duff, B.~E.~W.~Nilsson, C.~N.~Pope and N.~P.~Warner,
  {\it On The Consistency Of The Kaluza-Klein Ansatz},
  Phys.\ Lett.\  B {\bf 149} (1984) 90.

\bibitem{DeWitNicolaiConsistency}
  B.~de Wit and H.~Nicolai,
  {\it The Consistency of the $S^7$ Truncation in D=11 Supergravity},
  Nucl.\ Phys.\  B {\bf 281} (1987) 211.

\bibitem{ConsistentSphere}
  M.~Cvetic, H.~Lu and C.~N.~Pope,
  {\it Consistent Kaluza-Klein sphere reductions},
  Phys.\ Rev.\  D {\bf 62} (2000) 064028
  [arXiv:hep-th/0003286].

\bibitem{DuffPopeConsistentKK}
  M.~J.~Duff and C.~N.~Pope,
  {\it Consistent Truncations In Kaluza-Klein Theories},
  Nucl.\ Phys.\  B {\bf 255} (1985) 355.

\bibitem{CoquereauxJadczykConsistency}
  R.~Coquereaux and A.~Jadczyk,
  {\it Consistency of the G invariant Kaluza-Klein scheme},
  Nucl.\ Phys.\  B {\bf 276} (1986) 617.

\bibitem{Chatzistavrakidis2007}
  A.~Chatzistavrakidis, P.~Manousselis, N.~Prezas and G.~Zoupanos,
  {\it On the consistency of coset space dimensional reduction},
  Phys.\ Lett.\  B {\bf 656} (2007) 152
  [arXiv:0708.3222 [hep-th]];
  {\it Coset Space Dimensional Reduction of Einstein--Yang--Mills theory},
  Fortsch.\ Phys.\  {\bf 56} (2008) 389
  [arXiv:0712.2717 [hep-th]].

\bibitem{HullReidEdwardsStringTwistedTori}
  C.~M.~Hull and R.~A.~Reid-Edwards,
  {\it Flux compactifications of string theory on twisted tori},
  arXiv:hep-th/0503114.

\bibitem{Dall'AgataPrezas-ScherkSchwarz}
  G.~Dall'Agata and N.~Prezas,
  {\it Scherk-Schwarz reduction of M-theory on G2-manifolds with fluxes},
  JHEP {\bf 0510} (2005) 103
  [arXiv:hep-th/0509052].

\bibitem{GauntlettVarela}
  J.~P.~Gauntlett and O.~Varela,
  {\it Consistent Kaluza-Klein Reductions for General Supersymmetric AdS
  Solutions},
  Phys.\ Rev.\  D {\bf 76}, 126007 (2007)
  [arXiv:0707.2315 [hep-th]].

\bibitem{GauntlettKimVarelaWaldram}
  J.~P.~Gauntlett, S.~Kim, O.~Varela and D.~Waldram,
  {\it Consistent supersymmetric Kaluza--Klein truncations with massive modes},
  arXiv:0901.0676 [hep-th].

\bibitem{GaugingHeisenberg}
  R.~D'Auria, S.~Ferrara, M.~Trigiante and S.~Vaula,
  {\it Gauging the Heisenberg algebra of special quaternionic manifolds},
  Phys.\ Lett.\  B {\bf 610} (2005) 147
  [arXiv:hep-th/0410290].

\bibitem{TomasMirrorSymFl}
  A.~Tomasiello,
  {\it Topological mirror symmetry with fluxes},
  JHEP {\bf 0506} (2005) 067
  [arXiv:hep-th/0502148].      
      
\bibitem{GLW1} M.~Grana, J.~Louis and D.~Waldram,
  {\it Hitchin functionals in $N = 2$ supergravity},
  JHEP {\bf 0601}, 008 (2006)
  [arXiv:hep-th/0505264].

\bibitem{BedulliVezzoni}
	L.~Bedulli and L.~Vezzoni
	{\it The Ricci tensor of SU(3)-manifolds},
	J.\ Geom.\ Phys.\ 57 (2007), n.$\:$4, 1125
	[arXiv:math/0606786].  

\bibitem{Michelson}
J.~Michelson,
  {\it Compactifications of type IIB strings to four dimensions with non-trivial
  classical potential},
  Nucl.\ Phys.\  B {\bf 495} (1997) 127
  [arXiv:hep-th/9610151].

\bibitem{N=2withTensor1} 
G.~Dall'Agata, R.~D'Auria, L.~Sommovigo and S.~Vaula,
  {\it $D = 4$, $N = 2$ gauged supergravity in the presence of tensor multiplets},
  Nucl.\ Phys.\  B {\bf 682}, 243 (2004)
  \hbox{[arXiv:hep-th/0312210]}.

\bibitem{N=2withTensor2}
R.~D'Auria, L.~Sommovigo and S.~Vaula,
  {\it $N = 2$ supergravity Lagrangian coupled to tensor multiplets with  electric
  and magnetic fluxes},
  JHEP {\bf 0411}, 028 (2004)
  \hbox{[arXiv:hep-th/0409097]}.

\bibitem{D'AuriaFerrFre}
R.~D'Auria, S.~Ferrara and P.~Fre,
  {\it Special and quaternionic isometries: general couplings in N=2 supergravity
  and the scalar potential},
  Nucl.\ Phys.\  B {\bf 359} (1991) 705.

\bibitem{FerraraSabharwal}
  S.~Ferrara and S.~Sabharwal,
  {\it Quaternionic Manifolds for Type II Superstring Vacua of Calabi-Yau
  Spaces},
  Nucl.\ Phys.\  B {\bf 332}, 317 (1990).

\bibitem{Calderbank}
  D.~M.~J.~Calderbank and H.~Pedersen,
  {\it Selfdual Einstein metrics with torus symmetry},
  arXiv:math/0105263.

\bibitem{Strominger}
  A.~Strominger,
  {\it Loop corrections to the universal hypermultiplet},
  Phys.\ Lett.\  B {\bf 421}, 139 (1998)
  [arXiv:hep-th/9706195].

\bibitem{AMTV}
  I.~Antoniadis, R.~Minasian, S.~Theisen and P.~Vanhove,
  {\it String loop corrections to the universal hypermultiplet},
  Class.\ Quant.\ Grav.\  {\bf 20}, 5079 (2003)
  [arXiv:hep-th/0307268].

\bibitem{AmirShamit}
S.~Kachru and A.~K.~Kashani-Poor,
  {\it Moduli potentials in type IIA compactifications with RR and NS flux},
  JHEP {\bf 0503} (2005) 066
  [arXiv:hep-th/0411279].

\bibitem{MaldacenaNunez}
  J.~M.~Maldacena and C.~Nunez,
  {\it Supergravity description of field theories on curved manifolds and a no  go theorem,}
  Int.\ J.\ Mod.\ Phys.\  A {\bf 16} (2001) 822
  [arXiv:hep-th/0007018].

\bibitem{LustMarchMartTsimpis}
  D.~Lust, F.~Marchesano, L.~Martucci and D.~Tsimpis,
  {\it Generalized non-supersymmetric flux vacua},
  JHEP {\bf 0811} (2008) 021
  [arXiv:0807.4540 [hep-th]].

\bibitem{KoerberMartucci}
  P.~Koerber and L.~Martucci,
  {\it From ten to four and back again: how to generalize the geometry},
  JHEP {\bf 0708}, 059 (2007)
  [arXiv:0707.1038 [hep-th]].

\bibitem{ShiuTUDouglas}
  G.~Shiu, G.~Torroba, B.~Underwood and M.~R.~Douglas,
  {\it Dynamics of Warped Flux Compactifications},
  JHEP {\bf 0806} (2008) 024
  [arXiv:0803.3068 [hep-th]].

\bibitem{WhatIsSpecialKaehler?}
B.~Craps, F.~Roose, W.~Troost and A.~Van Proeyen,
  {\it What is special K\"ahler geometry?},
  Nucl.\ Phys.\  B {\bf 503} (1997) 565
  [arXiv:hep-th/9703082].



\end{thebibliography}
\end{document}